\documentclass[fleqn, usenatbib]{mnras}

\usepackage{newtxtext, newtxmath}

\usepackage[T1]{fontenc}

\usepackage{bm}

\DeclareRobustCommand{\VAN}[3]{#2}
\let\VANthebibliography\thebibliography
\def\thebibliography{\DeclareRobustCommand{\VAN}[3]{##3}\VANthebibliography}

\usepackage{graphicx}	
\usepackage{amsmath}	
\usepackage{xcolor}     

\usepackage{orcidlink}

\usepackage{multirow}
\usepackage{threeparttable}
\usepackage{soul}

\usepackage{enumitem}

\newcommand{\LISA}{\emph{LISA}}
\newcommand{\BALROG}{\textsc{balrog}}

\title[LISA verification binaries]{
Identifying LISA verification binaries among the Galactic population of double white dwarfs
}

\author[E. Finch et al.]{Eliot Finch$^{\orcidlink{0000-0002-1993-4263}}$,$^{1}$\thanks{E-mail: efinch@star.sr.bham.ac.uk}
Giorgia Bartolucci,$^{1}$
Daniel Chucherko,$^{1}$
Ben G. Patterson,$^{1}$
Valeriya Korol$^{\orcidlink{0000-0002-6725-5935}}$,$^{1,2}$ 
\newauthor
Antoine Klein$^{\orcidlink{0000-0001-5438-9152}}$,$^{1}$
Diganta Bandopadhyay$^{\orcidlink{0000-0003-0975-5613}}$,$^{1}$
Hannah Middleton$^{\orcidlink{0000-0001-5532-3622}}$,$^{1}$
Christopher J. Moore$^{\orcidlink{0000-0002-2527-0213}}$,$^{1}$ and
\newauthor
Alberto Vecchio$^{\orcidlink{0000-0002-6254-1617}}$$^{1}$
\\
$^{1}$Institute for Gravitational Wave Astronomy \& School of Physics and Astronomy, University of Birmingham, Birmingham, B15 2TT, UK \\
$^{2}$Max-Planck-Institut f\"{u}r Astrophysik, Karl-Schwarzschild-Stra\ss e 1, 85741 Garching, Germany
}

\date{Accepted XXX. Received YYY; in original form ZZZ}

\pubyear{2022}

\begin{document}
\label{firstpage}
\pagerange{\pageref{firstpage}--\pageref{lastpage}}
\maketitle

\begin{abstract}
    Double white dwarfs (DWDs) will be the most numerous gravitational-wave (GW) sources for the Laser Interferometer Space Antenna (\LISA{}). 
    Most of the Galactic DWDs will be unresolved and will superpose to form a confusion noise foreground, the dominant \LISA{} noise source around $\sim 0.5\text{--}3\,\mathrm{mHz}$. 
    A small fraction of these sources will stand out from the background and be individually detectable.
    Uniquely among GW sources, a handful of these binaries will be known in advance from electromagnetic (EM) observations and will be guaranteed sources of detectable GWs in the \LISA{} band; these are known as verification binaries (VBs). 
    High-cadence photometric surveys are continuously discovering new VB systems, and their number will continue to grow ahead of the launch of \LISA{}. 
    We analyse, in a fully Bayesian framework, all the currently known VB candidates with the latest design requirements for the \LISA{} mission and find that 25 of the considered sources can be detected within a $4\,\mathrm{yr}$ observation time. 
    We explore what can be expected from GW observations, both alone and in combination with EM observations, and estimate the VB's time to detection in the early months of \LISA{} operations.
    We also show how VBs can be analysed in the case where their GW signals compete with many other unknown binary signals (both resolved and unresolved) from a realistic Galactic population of DWDs.
\end{abstract}




\section{Introduction}\label{sec:introduction}

The Laser Interferometer Space Antenna (\LISA{}) is a gravitational-wave (GW) observatory currently under development for science operations in the 2030s~\citep{2017arXiv170200786A}. 
The \LISA{} design is optimised for sensitivity to GWs in the mHz range and the instrument will provide the first look at the GW sky in the frequency band $\sim 0.1 \text{--} 500\,\mathrm{mHz}$. 

From the first inception of a mission concept aimed at the mHz GW spectrum~\citep{LISA-proposal-1993}, it was realised that the Galactic population (GP) of short-period ($\lesssim 1\,\mathrm{h}$) ultra-compact binaries (UCBs) -- white dwarfs, neutron stars, and stellar-mass black holes -- represented a copious reservoir of detectable GW sources~\citep{1987A&A...176L...1L, 1990ApJ...360...75H}. 
It is now clear that at \LISA{}'s requirement sensitivity~\citep{2021arXiv210801167B} the instrument will be able to individually resolve tens of thousands of these UCBs~\citep{LISA:2022yao}. 
Galactic double white dwarfs (DWDs), both detached and interacting, will constitute the overwhelming majority of detected UCBs~\citep{Nelemans:2003ha, Nissanke:2012eh, Korol:2017qcx, Korol:2021pun, Kremer:2017xrg, Lamberts:2019nyk, Breivik:2019lmt, 2020ApJ...893....2L}. 
\LISA{} will also be sensitive enough to observe several tens to hundreds of DWDs harboured in Milky Way satellite galaxies~\citep{Korol:2018ulo, Korol:2020lpq, Roebber:2020hso}. 
UCBs with neutron stars and/or stellar-mass black holes are also expected to be detected, although in much smaller numbers~\citep{1990ApJ...360...75H, Nelemans:2001hp, Lamberts:2018cge, Andrews:2019plw, Lau:2019wzw, Wagg:2021cst}. 

Since the population of short-period UCBs is so abundant, the systems that \LISA{} can resolve will just be the `tip of the iceberg'. 
The incoherent superposition of GWs from the remaining unresolved sources in this population will produce a stochastic foreground signal, known as a confusion noise~\citep{1990ApJ...360...75H, Ruiter:2007xx, Farmer:2003pa, Georgousi:2022uyt}.
This confusion noise will actually dominate over \LISA{}'s instrumental noise in the frequency region $\sim 0.5\text{--}3\,\mathrm{mHz}$.

\begin{figure*}
    \centering
    \includegraphics[width=\textwidth]{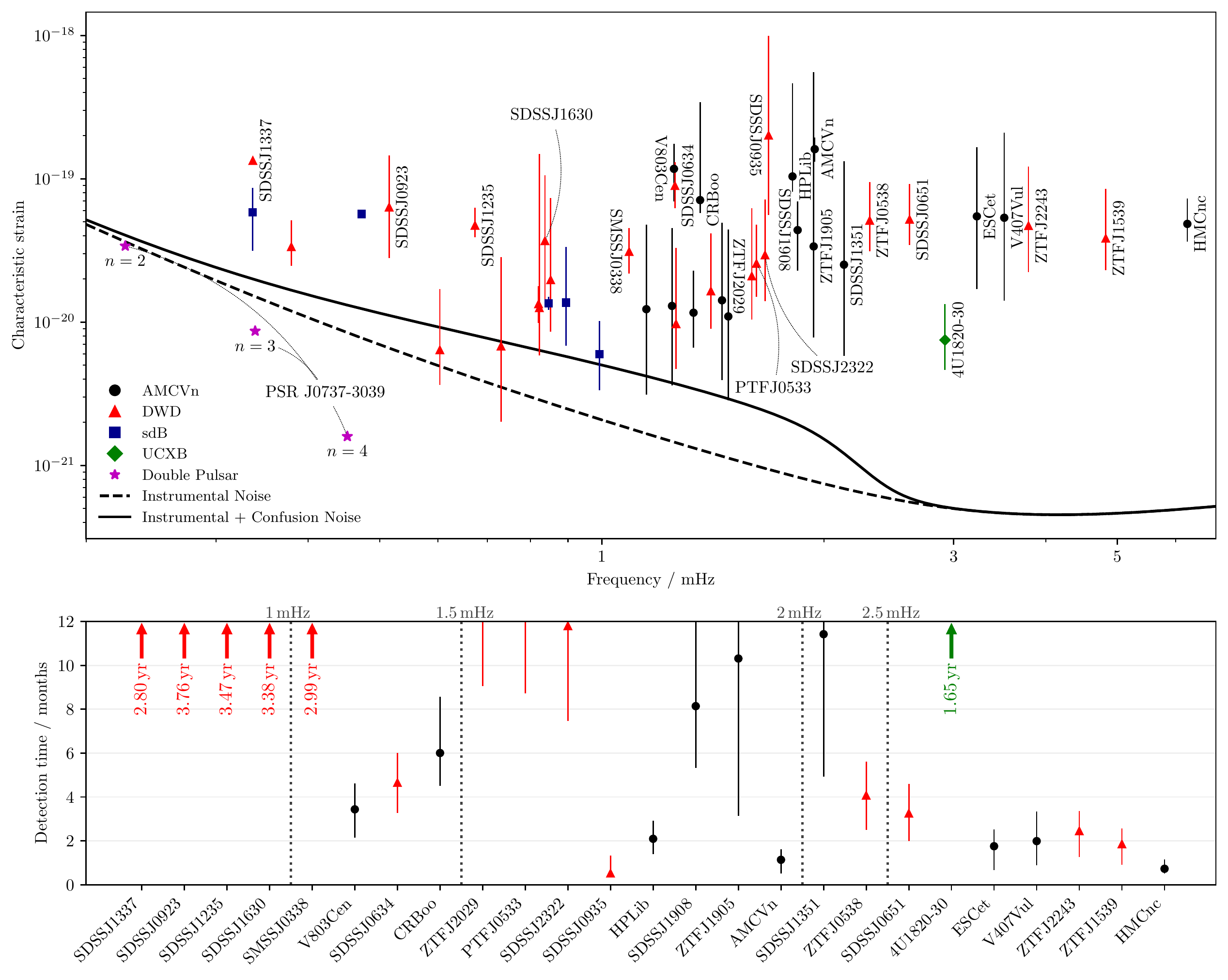}
    \caption{
    \textit{Top:} The characteristic strain, $h_c$, of all 43 VB candidates considered in this study (and the double pulsar) .
    This is compared to the \LISA{} instrumental noise $\sqrt{f S_{\rm inst}}$ and to the total noise including the foreground confusion $\sqrt{f (S_{\rm inst}+S_{\rm conf})}$ for a mission duration of $T_{\rm obs} = 4\,\mathrm{yr}$. 
    The ratio $h_c/\sqrt{f (S_{\rm inst}+S_{\rm conf})}$ gives the SNR of the source. 
    Here, we use an SNR calculated in the low-frequency limit, averaged over \LISA{} orbital modulations and polarisation angle; further details can be found in Appendix.~\ref{sec:noisecurves}. 
    Error bars indicate the full range of possible $h_c$ given the measurement uncertainties in VB component masses, distances, inclinations, and frequencies (these sources of uncertainty being in decreasing order of significance).
    Also shown are the first few harmonics of the double pulsar PSR J0737-3039 that is on an eccentric orbit with a period of $2.45\,\mathrm{h}$; although it is close, this is not expected to be detectable with \LISA{} (see Section~\ref{subsec:double_pulsar}). 
    \textit{Bottom:} Time to detection for the 25 VBs that satisfy $\mathrm{SNR}_{4\mathrm{yr}} > 6$ (i.e., the 25 sources that are detectable within a $4\,\mathrm{yr}$ mission).
    Markers and error bars show the median and 50\% error bars on the detection time, including both the astrophysical uncertainty and the \LISA{} orbital uncertainty as described in Section~\ref{sec:LISA_early}.
    Sources are shown in order of increasing frequency from left to right.
    Sources that will not be detectable in the first year of \LISA{} science operations are indicated with arrows, and their approximate detection time is shown. 
    We emphasise that no VB sources with frequencies $f_0<1 \mathrm{mHz}$ are expected to be detectable within the first year of \LISA{} operations.
    }
    \label{fig:CharacteristicStrain}
\end{figure*}

Uniquely among GW sources, DWDs are also bright and persistent sources of electromagnetic (EM) radiation~\citep[however, we note that the identification of EM emission from massive black hole binaries might also be expected by the time \LISA{} flies;][]{Xin:2021mmk}. 
This means that in some cases they can be detected and studied before the launch of \LISA{}.
In the early 1990s, a few UCBs were already known to be guaranteed GW sources for \LISA{}~\citep[see, for example,][and references therein]{LISA-proposal-1993, LISA-Pre-Phase-A}. 
Known UCBs whose radiation will be detectable by future GW missions such as \LISA{} are called verification binaries (VBs)~\citep[this idea goes back to, e.g.,][]{VB-Phinney:2001}.
VBs offer a guaranteed detection of GWs within the first few weeks of the \LISA{} mission and will be useful 
for testing the \LISA{} instrument and maximising its scientific output. They also offer a new opportunity to study the astrophysics of compact binaries using both their GW and EM emission~\citep[e.g.][]{Marsh:2011yj}.

The list of known VBs has grown over the years~\citep{Stroeer:2006rx,Kupfer:2018jee,2023arXiv230212719K}.
New and interesting systems are continuously being discovered by surveys such as the Extremely Low Mass Survey~\citep{Brown:2010sa, 2020ApJ...889...49B}, ATLAS~\citep{2018PASP..130f4505T}, OmegaWhite~\citep{2015MNRAS.454..507M}, {\it Gaia}~\citep{Gaia:2016zol} and the Zwicky Transient Facility~\citep[ZTF;][]{2019PASP..131a8002B, Graham:2019qsw}.
Their number will increase further with new surveys such as SDSS-V~\citep{2017arXiv171103234K}, 
BlackGEM~\citep{2015ASPC..496..254B}, the Gravitational-wave Optical Transient Observer~\citep{2017NatAs...1..741S},
and the Vera Rubin Observatory~\citep{LSST:2008ijt}. 
It is reasonable to expect that the number of known VBs may reach $\mathcal{O}(10^3)$ by the time \LISA{} flies~\citep[e.g.\ ][]{Korol:2017qcx}.

An early study of how well \LISA{} could measure VB properties was performed by \citet{Stroeer:2006rx}, who identified approximately eight VBs above a signal-to-noise ratio (SNR) detection threshold of five and that are therefore expected to be resolvable.
Later, \citet{Kupfer:2018jee} used updated EM observations to identify 16 VBs above a similar SNR threshold.
Both of these studies used a Fisher matrix formalism to estimate the measurement uncertainties, and both studies used a simplified model for the \LISA{} response to GWs involving just two noise-orthogonal channels.

This work represents an update to these previous studies on VBs, making use of the most recent version of the detectable Galactic binaries table from \citet{KupferTable, 2023arXiv230212719K}.
We also use a realistic model for the \LISA{} response to GWs, incorporating all three time-delay interferometry (TDI) output channels and an instrumental noise curve associated with the latest design requirements for the \LISA{} mission~\citep{2021arXiv210801167B}.
Details of all 43 VB candidates considered in this study are summarised in Table~\ref{tab:EM_table}, and the top panel of Fig.~\ref{fig:CharacteristicStrain} illustrates the characteristic strain of this set of VBs alongside the chosen \LISA{} noise curve.
Note that we refer to the full set of binaries considered as ``VB candidates'', since not all of them will be detectable in \LISA{}.
Those sources that satisfy the detection criteria are then known as VBs.

The scope of the study presented here is quite broad, and to help guide the reader we present a summary of the key sections below:
\begin{itemize}[leftmargin=0.5cm, labelwidth=0.5cm]
    \item[{\textbullet}] Section~\ref{sec:VBstudy}, \emph{``Updated VB parameter estimation study''}, contains an updated parameter estimation study on how well the VB properties can be measured from GW observations alone. In this section each VB is considered in isolation and the instrumental noise properties are assumed to be known perfectly. We also study how the SNR for each VB accumulates over time, with a particular focus on the early months of the \LISA{} mission so we can estimate the time to detection of the loudest VBs.
    \vspace{0.2cm}
    \item[{\textbullet}] Section~\ref{sec:EMpriors}, \emph{``EM -- GW synergies''}, investigates the effect of incorporating EM priors in the GW analysis to address how the accuracy of the measurements can be improved~\citep[this complements][who considered the influence that \LISA{} measurements will have on the analysis of EM observations]{2021arXiv211200145J}.
    \vspace{0.2cm}
    \item[{\textbullet}] Section~\ref{sec:unknown_noise}, \emph{``Accounting for unknown noise levels''}, reperforms the VB parameter estimation study, but now treats the noise in each \LISA{} TDI channel as an unknown parameter to be inferred from the data simultaneously with the VB properties. This shows that the results in Section~\ref{sec:VBstudy} are robust to the addition of uncertain levels of instrumental or confusion noise.
    \vspace{0.2cm}
    \item[{\textbullet}] Section~\ref{sec:confusion}, \emph{``Accounting for source confusion''}, performs an analysis of a single typical VB while including all other DWD signals from a realistic Galaxy realisation (predicted from population synthesis). This involves modelling above-threshold (but EM-dark) sources nearby in frequency to the VB, while simultaneously accounting for the below-threshold sources (which constitute the confusion foreground) via a variable noise level.
\end{itemize}

\section{Updated Verification Binary Parameter Estimation Study}\label{sec:VBstudy}

Table~\ref{tab:EM_table} summarises the properties of the currently known candidate VBs split into four sub categories: detached DWDs, accreting DWDs (also known as AM CVns), hot subdwarfs with a white dwarf companion (sdBs), and ultra-compact X-ray binaries (UCXBs). 
The orbital periods of these binaries are typically well determined from the source variability via photometry or spectroscopy.
Component masses are harder to measure because of the intrinsic faintness and the compact configuration~\citep[e.g., see][]{2018MNRAS.480.4505J,Rebassa-Mansergas:2018mtp}.
It is especially difficult for AM CVn type sources because only the accretion disc and, more exceptionally, the accretor are visible in the spectra. 
Typically, both component masses can be estimated for eclipsing systems.
Distance is yet another parameter difficult to derive from EM observations. 
For many VBs, distance could be derived thanks to the arrival of the {\it Gaia} data, which was considered in \citet{Kupfer:2018jee}. 
Here we use up-to-date distance estimates derived in \citet{2023arXiv230212719K} based on the latest {\it Gaia} data release 3. 
We note, however, that the highest frequency binary (HMCnc) -- expected to be the `loudest' \LISA{} source among currently known Galactic binaries -- in the sample is lacking parallax measurement; as a consequence, its distance remains highly uncertain. 
As in \citet{2023arXiv230212719K}, for HMCnc we consider a range of possible distances between 5 and 10\,kpc, which comprises various estimates in the literature.
Finally, we note that the inclination is a degenerate parameter that can generally only be well constrained for eclipsing or nearly eclipsing systems.

\begin{table*}
\centering
\caption{
Source properties measured from EM observations for the set of known VBs (accessed on 2022 September 2) from~\citet{KupferTable}~\citep[see also][]{2023arXiv230212719K}.
Masses, distances, and inclination angles with no available uncertainty information are quoted to two significant figures, and values stated in square brackets represent indirect estimates based on theoretical arguments~\citep{Kupfer:2018jee}. 
We note that HMCnc has an essentially unconstrained distance, and in the parameter estimation study in Section~\ref{sec:VBstudy} below we adopt the median (7500\,pc) of the stated range as our fiducial value; we note that this differs from the value used in 
\citet{Kupfer:2018jee}, although see the discussion in Section~5 of that paper. 
Inclinations are only in the range $0\text{--}90\,\mathrm{deg}$ due to the inability of EM observations to distinguish between an inclination of $x$ degrees and $(180-x)$ degrees. GW observations from \LISA{} will resolve this degeneracy and find the true inclination of each system in the range $0\text{--}180\,\mathrm{deg}$. 
(Note that when injecting VBs we choose to inject with the inclination given in the table.) 
Eclipsing systems are denoted with $\ast$.}
\label{tab:EM_table}
\begin{threeparttable} 
 \begin{tabular}{ccccccccc}
\hline
 \textbf{Type} & \multirow{2}{*}{$l$ / deg} & \multirow{2}{*}{$b$ / deg} & \multirow{2}{*}{$P$ / s} & \multirow{2}{*}{$m_1$ / $M_{\odot}$} & \multirow{2}{*}{$m_2$ / $M_{\odot}$} & \multirow{2}{*}{D / pc} &\multirow{2}{*}{$\iota$ / deg} & \multirow{2}{*}{Refs.} \\ Source &&&&&&& \\
\hline
\textbf{AMCVn} &&&&&& & &\\
HMCnc& 120.4387 &-4.7040 &321.5291290(10) & 0.55 & 0.27 &[5000 \text{--} 10000] &38 &1, 2, 3 \\
V407Vul& -65.0093 &46.7833 &569.39623(13) &[0.80 $\pm$ 0.10] &[0.177 $\pm$ 0.071] &2090 $\pm$ 680 &[60] &2 \\
ESCet $\ast$ &24.6080 & -20.3339 &620.21125(30) &[0.80 $\pm$ 0.10] &[0.161 $\pm$ 0.064] &1780 $\pm$ 230 &[60] &4 \\
SDSSJ1351&-151.6161 & 4.4721 &939.0 $\pm$ 7.2 &[0.80 $\pm$ 0.10] &[0.100 $\pm$ 0.040] &1530 $\pm$ 760 &[60] &5 \\
AMCVn& 170.3818 &37.4426 & 1028.73220(30) &0.680 $\pm$ 0.060 &0.125 $\pm$ 0.012 & 302.0 $\pm$ 3.0 &43.0 $\pm$ 2.0 &6 \\
SDSSJ1908& -61.7867 &61.4542 &1085.1080(10) &[0.80 $\pm$ 0.10] &[0.085 $\pm$ 0.034] &977 $\pm$ 32 &15.0 $\pm$ 5.0 & 7, 8 \\
HPLib&-124.9155 & 4.9599 & 1102.700(50) &0.65 $\pm$ 0.15 &0.07 $\pm$ 0.20 & 280.0 $\pm$ 3.0 &30.0 $\pm$ 4.0 &9, 10 \\
PTF1919 $\ast$ & -51.0016 &69.0291 & 1347.354(20) &[0.80 $\pm$ 0.10] &[0.066 $\pm$ 0.026] &1360 $\pm$ 470 &[60] & 11 \\
 CX1751& -91.9424 &-6.2528 &1374.00(60) &[0.80 $\pm$ 0.10] &[0.064 $\pm$ 0.026] &1130 $\pm$ 260 &[60] & 12 \\
CRBoo&-157.7309 &17.8971 & 1471.306(50) &0.89 $\pm$ 0.21 &0.07 $\pm$ 0.22 & 351.0 $\pm$ 5.0 &30 & 10, 13 \\
V803Cen&-143.8365 & -30.3168 & 1596.4 $\pm$ 1.2 &0.97 $\pm$ 0.20 &0.084 $\pm$ 0.025 & 287.0 $\pm$ 5.0 &13.5 $\pm$ 1.5 & 10, 14 \\
KLDra& -25.8709 &78.3217 & 1501.806(30) & 0.76 &0.057 &930 $\pm$ 91 &[60] & 15 \\
PTF0719& 104.3844 &26.5213 & 1606.2 $\pm$ 1.2 &[0.80 $\pm$ 0.10] &[0.053 $\pm$ 0.021] & 840 $\pm$ 200 &[60] & 16 \\
CPEri&42.1289 & -26.4276 & 1740(84) &[0.80 $\pm$ 0.10] &[0.049 $\pm$ 0.020] & 750 $\pm$ 200 &[60] & 17 \\
 ZTFJ1905 $\ast$ & -66.2000 &53.6764 & 1032.16441(62) &[0.80 $\pm$ 0.10] &[0.090 $\pm$ 0.035] & 700 $\pm$ 600 & 70 $\pm$ 20 & 18 \\
\textbf{DWD} &&&&&& & &\\
SDSSJ0651 $\ast$ & 101.3338 & 5.8064 & 765.206543(55) &0.247 $\pm$ 0.013 &0.490 $\pm$ 0.020 & 960 $\pm$ 370 & $86.9\substack{+1.6 \\ -1.0}$ & 19, 20 \\
SDSSJ0935& 130.9744 &28.0938 & 1188(42) &0.312 $\pm$ 0.019 &0.75 $\pm$ 0.24 & 400 $\pm$ 200 &[60] & 21, 22 \\
SDSSJ1630&-128.2284 &63.0527 & 2388.0 $\pm$ 6.0 &0.298 $\pm$ 0.019 &0.76 $\pm$ 0.24 & 850 $\pm$ 170 &[60] & 21, 23 \\
SDSSJ0923& 133.7104 &14.4288 & 3884(43) &0.275 $\pm$ 0.011 &0.76 $\pm$ 0.23 & 288.0 $\pm$ 5.0 &[60] & 21, 24 \\
 ZTFJ1539 $\ast$ &-154.9724 &66.1616 &414.7915404(29) & $0.610\substack{+0.017 \\ -0.022}$ &0.210 $\pm$ 0.015 & 2500 $\pm$ 1300 &$84.15\substack{+0.64 \\ -0.57}$ & 18 \\
 ZTFJ0538 $\ast$ &84.8261 &-3.4567 &866.60331(16) &0.450 $\pm$ 0.050 &0.320 $\pm$ 0.030 &1000 $\pm$ 370 & $85.430\substack{+0.070 \\ -0.090}$ & 18 \\
 PTFJ0533&82.9058 & -21.1234 & 1233.97298(17) & $0.652\substack{+0.037 \\ -0.040}$ &0.167 $\pm$ 0.030 &1170 $\pm$ 390 & $72.8\substack{+0.8 \\ -1.4}$ & 18 \\
 ZTFJ2029 $\ast$ & -45.5630 &33.4339 &1252.056499(41) &0.320 $\pm$ 0.040 &0.300 $\pm$ 0.040 &1100 $\pm$ 640 &$86.64\substack{+0.70 \\ -0.40}$ & 18 \\
 ZTFJ1749 $\ast$ & -92.9622 &32.8224 & 1586.03389(44) & $0.400\substack{+0.070 \\ -0.050}$ & $0.280\substack{+0.050 \\ -0.040}$ & 2000 $\pm$ 1200 & $85.5\substack{+1.4 \\ -1.1}$ & 18 \\
 ZTFJ2243 $\ast$ &13.2384 &53.9599 & 527.934890(32) & $0.349\substack{+0.093 \\ -0.074}$ &$0.38\substack{+0.11 \\ -0.07}$ &1760 $\pm$ 730 & $81.9\substack{+1.3 \\ -0.7}$ & 25 \\
SDSSJ2322&-6.5666 & 8.4572 & 1201.4 $\pm$ 5.9 &0.340 $\pm$ 0.020 &$>$0.17 & 860 $\pm$ 210 &[60] & 26 \\
SDSSJ1235&-178.2132 &17.9524 & 2970.4 $\pm$ 4.3 &0.350 $\pm$ 0.010 & $0.270\substack{+0.060 \\ -0.020}$ &446 $\pm$ 28 &27.0 $\pm$ 3.8 & 27 \\
 ZTFJ0722& 115.8862 & -40.2651 &1422.548655(71) &0.380 $\pm$ 0.040 &0.330 $\pm$ 0.030 &1460 $\pm$ 780 &89.66 $\pm$ 0.22 & 18 \\
 ZTFJ1901& -53.1907 &74.6334 & 2436.10817(93) &0.360 $\pm$ 0.040 &0.360 $\pm$ 0.050 &909 $\pm$ 78 &87.28 $\pm$ 0.50 & 18 \\
SMSSJ0338&80.4851 &59.4015 & 1836(32) &0.230 $\pm$ 0.015 & $0.380\substack{+0.050 \\ -0.030}$ &536 $\pm$ 16 &69.0 $\pm$ 9.0 & 28 \\
SDSSJ0634&97.0793 &14.8391 & 1591(29) & $0.452\substack{+0.070 \\ -0.062}$ & $0.209\substack{+0.034 \\ -0.021}$ &433 $\pm$ 16 &37.0 $\pm$ 7.0 & 28 \\
SDSSJ1337&-177.1107 &45.5716 &5942.95(30) &0.510 $\pm$ 0.010 &0.320 $\pm$ 0.010 & 113.78 $\pm$ 0.57 &13.0 $\pm$ 1.0 & 29 \\
 ZTFJ2320& 8.7132 &38.0937 &3314.7998(40) &0.690 $\pm$ 0.030 &0.200 $\pm$ 0.010 &1480 $\pm$ 860 & $84.5\substack{+2.7 \\ -3.2}$ & 18 \\
SDSSJ1043& 160.1506 &-2.0480 & 2739(79) &0.183 $\pm$ 0.010 &0.76 $\pm$ 0.25 & 2800 $\pm$ 1200 &[60] & 30 \\
SDSSJ0822 $\ast$ & 120.6776 &11.0965 & 2430.07250(10) &0.304 $\pm$ 0.014 &0.524 $\pm$ 0.050 & 1300 $\pm$ 1200 &87.70 $\pm$ 0.20 & 31 \\
SDSSJ0106&11.4543 & -15.7928 & 2345.8 $\pm$ 1.7 &0.188 $\pm$ 0.011 &$0.57\substack{+0.22 \\ -0.07}$ & 820 $\pm$ 440 & 67 $\pm$ 13 & 32 \\
 WD0957&-151.4766 & -67.3014 &5269.810804(73) &0.370 $\pm$ 0.020 &0.320 $\pm$ 0.030 & 163.70 $\pm$ 0.80 & 75 $\pm$ 15 & 33 \\
\textbf{sdB} &&&&&& & &\\
CDm30 $\ast$ &-138.8255 & -16.6150 & 4231.79186(15) &0.540 $\pm$ 0.020 &0.790 $\pm$ 0.010 & 355.0 $\pm$ 7.0 &82.900 $\pm$ 0.040 & 34 \\
 ZTFJ2130 $\ast$ & -11.8355 &54.4443 &2360.4062(14) &0.545 $\pm$ 0.020 &0.337 $\pm$ 0.015 & 1307 $\pm$ 42 &86.4 $\pm$ 1.0 & 18 \\
 HD265435  & 101.3348 &10.1443 & 5945.91743(28) &$0.63\substack{+0.13 \\ -0.12}$ &1.01 $\pm$ 0.15 &461 $\pm$ 12 &$64\substack{+14 \\ -5}$ & 35 \\
 ZTFJ1946& -52.0264 &52.0541 & 2013.82141(75) & $0.272\substack{+0.046 \\ -0.043}$ & $0.307\substack{+0.097 \\ -0.085}$ &2120 $\pm$ 300 & $77.1\substack{+1.6 \\ -1.2}$ & 18 \\
 ZTFJ0640&99.6393 &-5.4567 &2236.0160(16) &$0.39\substack{+0.12 \\ -0.09}$ & $0.325\substack{+0.030 \\ -0.015}$ &1580 $\pm$ 620 &65.3 $\pm$ 5.1 & 18 \\
 \textbf{UCXB} &&&&&& & &\\
4U1820-30& -84.8589 &-7.0267 &685.0 $\pm$ 4.0 &[1.4] &[0.069] &7600 &[60] & 36 \\
\hline
\end{tabular}
 \begin{tablenotes} 
 \small 
 \item {[1]~\citet{Strohmayer:2005uc}, [2]~\citet{Barros:2006fa}, [3]~\citet{Roelofs:2010uv}, [4]~\citet{Espaillat:2004ua}, [5]~\citet{Green:2018wnq}, [6]~\citet{Skillman:1999rpz}, [7]~\citet{Fontaine:2011hoj}, [8]~\citet{Kupfer:2015lqa}, [9]~\citet{Patterson:2002pce}, [10]~\citet{Roelofs:2007ny}, [11]~\citet{Levitan:2014qha}, [12]~\citet{Wevers:2016vdt}, [13]~\citet{Provencal:1997mpz}, [14]~\citet{Roelofs:2007nz}, [15]~\citet{Wood:2002wpm}, [16]~\citet{Levitan:2013tnb}, [17]~\citet{Howell:1991dfi}, [18]~\citet{Burdge:2020end}, [19]~\citet{Brown:2011gq}, [20]~\citet{Hermes:2012us}, [21]~\citet{Brown:2016fyb}, [22]~\citet{Kilic:2014exv}, [23]~\citet{Kilic:2011ej}, [24]~\citet{Brown:2010sa}, [25]~\citet{Burdge:2020bul}, [26]~\citet{Brown:2020uvh}, [27]~\citet{Kilic:2017uld}, [28]~\citet{Kilic:2021dtv}, [29]~\citet{Chandra:2021piy}, [30]~\citet{Brown:2017nkc}, [31]~\citet{Kosakowski:2021nqm}, [32]~\citet{Kilic:2011hn}, [33]~\citet{Moran:1997zk}, [34]~\citet{Geier:2013koq}, [35]~\citet{Pelisoli:2021ykn}, [36]~\citet{Chen:2020wan}}
 \end{tablenotes} 
 \end{threeparttable}
\end{table*}

\begin{table*}
\centering
\caption{GW-derived parameter estimates for the set of known VBs, with an integration time of 4 yr and an $\mathrm{SNR}_{4\mathrm{yr}} > 6$ detection threshold. 
For the remaining sources that do not satisfy this detection threshold, we do not report parameter estimates, since they are found to be uninformative. 
The values $\Delta \mathcal{A} / \mathcal{A}$, $f_0$, $\dot{f}$, and $\iota$ are given to a $1\sigma$ confidence, and $\Omega_{90}$ is the 90\% credible region sky localisation.}
\label{tab:GW_table}
\begin{tabular}{cccccccccc}
\hline
 \textbf{Type} & \multicolumn{3}{c}{SNR} & \multirow{2}{*}{$\frac{\Delta \mathcal{A}}{\mathcal{A}}$} & \multirow{2}{*}{$f_0$ / mHz} & \multirow{2}{*}{$\dot{f}$ / nHz yr$^{-1}$} & \multirow{2}{*}{$\iota$ / deg} & \multirow{2}{*}{$\Omega_{90}$ / deg$^2$} \\ Source & \color{gray} 1 yr & 4 yr & \color{gray} 10 yr &&&&&& \\
\hline
\textbf{AMCVn} & &&&&&& &\\
 HMCnc &\color{gray}49.1 & 98.1 &\color{gray}155 &0.086 & 6.22027624(18) & 23.579 $\pm$ 0.087 & 30 $\pm$ 10 & 0.26 \\
 V407Vul &\color{gray}46.4 &116 &\color{gray}183 &0.021 & 3.51249253(15) &2.602 $\pm$ 0.079 &59.9 $\pm$ 1.0 &0.097 \\
 ESCet &\color{gray}35.8 &115 &\color{gray}182 &0.021 & 3.22470771(14) &1.747 $\pm$ 0.072 &60.1 $\pm$ 1.0 & 0.21 \\
 SDSSJ1351 &\color{gray}5.39 & 20.5 & \color{gray}55.9 & 0.18 & 2.12992548(84) &0.23 $\pm$ 0.42 & 54 $\pm$ 14 & 27 \\
 AMCVn &\color{gray}29.9 & 94.9 &\color{gray}288 &0.095 & 1.94414061(18) &0.171 $\pm$ 0.084 & 38 $\pm$ 10 & 0.53 \\
 SDSSJ1908 &\color{gray}7.68 & 22.7 & \color{gray}67.2 & 0.15 & 1.84313450(79) &0.12 $\pm$ 0.40 & 37 $\pm$ 13 &7.1 \\
 HPLib &\color{gray}17.8 & 51.7 &\color{gray}151 & 0.10 & 1.81372994(32) &0.08 $\pm$ 0.16 & 31 $\pm$ 11 &4.9 \\
 PTF1919 &\color{gray}1.49 & 3.85 & \color{gray}8.18 &- &- &- & - &- \\
CX1751 &\color{gray}1.91 & 4.89 & \color{gray}10.2 &- &- &- & - &- \\
 CRBoo &\color{gray}8.82 & 22.0 & \color{gray}42.9 & 0.15 & 1.35933688(76) &0.05 $\pm$ 0.38 & 38 $\pm$ 13 & 28 \\
 V803Cen &\color{gray}13.4 & 32.7 & \color{gray}60.4 & 0.13 & 1.25281882(57) &0.04 $\pm$ 0.27 & 34 $\pm$ 12 & 11 \\
 KLDra &\color{gray}1.41 & 3.51 & \color{gray}6.73 &- &- &- & - &- \\
 PTF0719 &\color{gray}1.48 & 3.61 & \color{gray}6.65 &- &- &- & - &- \\
 CPEri &\color{gray}1.30 & 3.09 & \color{gray}5.54 &- &- &- & - &- \\
ZTFJ1905 &\color{gray}6.28 & 19.8 & \color{gray}60.0 &0.085 & 1.93767579(90) &0.16 $\pm$ 0.46 &69.4 $\pm$ 3.4 &9.7 \\
\textbf{DWD} & &&&&&& &\\
 SDSSJ0651 &\color{gray}15.9 & 84.0 &\color{gray}143 &0.012 & 2.61367341(22) &0.83 $\pm$ 0.10 &86.90 $\pm$ 0.33 &1.1 \\
 SDSSJ0935 &\color{gray}31.2 & 85.5 &\color{gray}222 &0.029 & 1.68350165(21) &0.296 $\pm$ 0.097 &59.8 $\pm$ 1.4 &1.0 \\
 SDSSJ1630 &\color{gray}2.86 & 6.39 & \color{gray}10.8 & 0.32 &0.8375210(31) &0.0 $\pm$ 2.3 & 52 $\pm$ 18 &470 \\
 SDSSJ0923 &\color{gray}3.06 & 6.42 & \color{gray}10.4 & 0.31 &0.5149755(33) &0.0 $\pm$ 1.8 & 52 $\pm$ 17 & 1193 \\
ZTFJ1539 &\color{gray}41.8 & 84.1 &\color{gray}133 &0.012 & 4.82169908(19) &8.005 $\pm$ 0.096 &84.15 $\pm$ 0.36 &0.073 \\
ZTFJ0538 &\color{gray}12.4 & 57.7 &\color{gray}128 &0.018 & 2.30786102(31) &0.62 $\pm$ 0.15 &85.43 $\pm$ 0.53 &3.2 \\
PTFJ0533 &\color{gray}3.78 & 10.1 & \color{gray}24.6 & 0.21 &1.6207811(18) &0.12 $\pm$ 0.91 & 70 $\pm$ 12 &104 \\
ZTFJ2029 &\color{gray}3.15 & 8.38 & \color{gray}19.9 & 0.14 &1.5973720(23) &0.1 $\pm$ 1.1 &86.2 $\pm$ 4.4 &116 \\
ZTFJ1749 &\color{gray}1.12 & 2.74 & \color{gray}5.09 &- &- &- & - &- \\
ZTFJ2243 &\color{gray}47.7 &104 &\color{gray}164 &0.010 & 3.78834595(18) &3.584 $\pm$ 0.088 &81.88 $\pm$ 0.32 &0.091 \\
 SDSSJ2322 &\color{gray}4.50 & 12.2 & \color{gray}31.2 & 0.24 &1.6647245(15) &0.09 $\pm$ 0.72 & 49 $\pm$ 16 & 93 \\
 SDSSJ1235 &\color{gray}2.99 & 6.46 & \color{gray}10.7 & 0.31 &0.6733028(29) &0.0 $\pm$ 1.7 & 51 $\pm$ 17 &960 \\
ZTFJ0722 &\color{gray}2.14 & 5.42 & \color{gray}10.9 &- &- &- & - &- \\
ZTFJ1901 &\color{gray}1.02 & 2.28 & \color{gray}3.84 &- &- &- & - &- \\
 SMSSJ0338 &\color{gray}3.08 & 7.26 & \color{gray}12.8 & 0.31 &1.0892653(30) &0.0 $\pm$ 1.4 & 57 $\pm$ 17 &222 \\
 SDSSJ0634 &\color{gray}10.2 & 25.0 & \color{gray}46.3 & 0.14 & 1.25675506(75) &0.04 $\pm$ 0.35 & 37 $\pm$ 13 & 25 \\
 SDSSJ1337 &\color{gray}3.61 & 7.35 & \color{gray}11.7 & 0.26 &0.3365330(24) &0.0 $\pm$ 1.2 & 49 $\pm$ 16 &793 \\
ZTFJ2320 & \color{gray}0.358 &0.765 & \color{gray}1.26 &- &- &- & - &- \\
 SDSSJ1043 & \color{gray}0.463 & 1.01 & \color{gray}1.69 &- &- &- & - &- \\
 SDSSJ0822 & \color{gray}0.995 & 2.22 & \color{gray}3.75 &- &- &- & - &- \\
 SDSSJ0106 &\color{gray}1.57 & 3.52 & \color{gray}5.96 &- &- &- & - &- \\
WD0957 &\color{gray}1.09 & 2.23 & \color{gray}3.58 &- &- &- & - &- \\
\textbf{sdB} & &&&&&& &\\
 CDm30 &\color{gray}2.50 & 5.21 & \color{gray}8.42 &- &- &- & - &- \\
ZTFJ2130 &\color{gray}1.06 & 2.39 & \color{gray}4.04 &- &- &- & - &- \\
HD265435 &\color{gray}1.59 & 3.24 & \color{gray}5.16 &- &- &- & - &- \\
ZTFJ1946 & \color{gray}0.541 & 1.25 & \color{gray}2.17 &- &- &- & - &- \\
ZTFJ0640 &\color{gray}1.12 & 2.54 & \color{gray}4.34 &- &- &- & - &- \\
 \textbf{UCXB} & &&&&&& &\\
 4U1820-30 &\color{gray}3.38 & 14.7 & \color{gray}23.6 & 0.21 &2.9197080(13) &0.78 $\pm$ 0.60 & 50 $\pm$ 15 & 34 \\
\hline
\end{tabular}
\end{table*}

\subsection{GW parameter estimation}\label{subsec:parameter_estimation}

We use the up-to-date candidate VB properties in Table~\ref{tab:EM_table} and a Bayesian inference pipeline to study the measurement of the VB source properties from their GW signals alone.
We emphasise that we do not use any prior EM-derived knowledge on the GW parameters in the analyses in this section; a study of the improvement obtained with a combined multimessenger analysis is left to Section~\ref{sec:EMpriors}.

Bayesian parameter estimation was performed using the \BALROG{} code~\citep{Roebber:2020hso, Buscicchio:2021dph, 2022arXiv220403423K}.
\BALROG{} simulates the \LISA{} mission, including DWD waveform generation and the \LISA{} response with mock noise, and has the capability to perform parameter estimation.
Unlike ground-based detectors, the \LISA{} data can be processed to produce three output channels (conventionally named $A$, $E$, and $T$) containing independent noise.
With the additional assumptions of stationarity and Gaussianity, the noise in each channel can be characterised by the power spectral density (PSD) and we write the likelihood as a product over the three independent channels as
\begin{align} \label{eq:log_like}
    P(d | h) \propto \prod_\alpha \exp\left(-\frac{1}{2}\langle d - h | d - h \rangle_\alpha \right),
\end{align}
where $h$ and $d$ are the frequency-domain representations of the signal model and observed data respectively in the three TDI channels: $\alpha=A,\;E,\;\mathrm{and}\;T$.
The inner product in each channel is defined as
\begin{equation} \label{eq:inner_prod}
    \langle a | b \rangle_\alpha = 2\sum_k \frac{a_k b^*_k + a^*_k b_k}{S_\alpha(f_k)} \delta f,
\end{equation}
where $S_\alpha(f_k)$ is the PSD in channel $\alpha$.
All quantities are in the frequency domain and the subscript $k$ indexes the frequency components in their discrete Fourier transforms. 
The frequency resolution is $\delta f=1/T_{\rm obs}$.

\begin{figure}
    \centering
    \includegraphics[width=\linewidth]{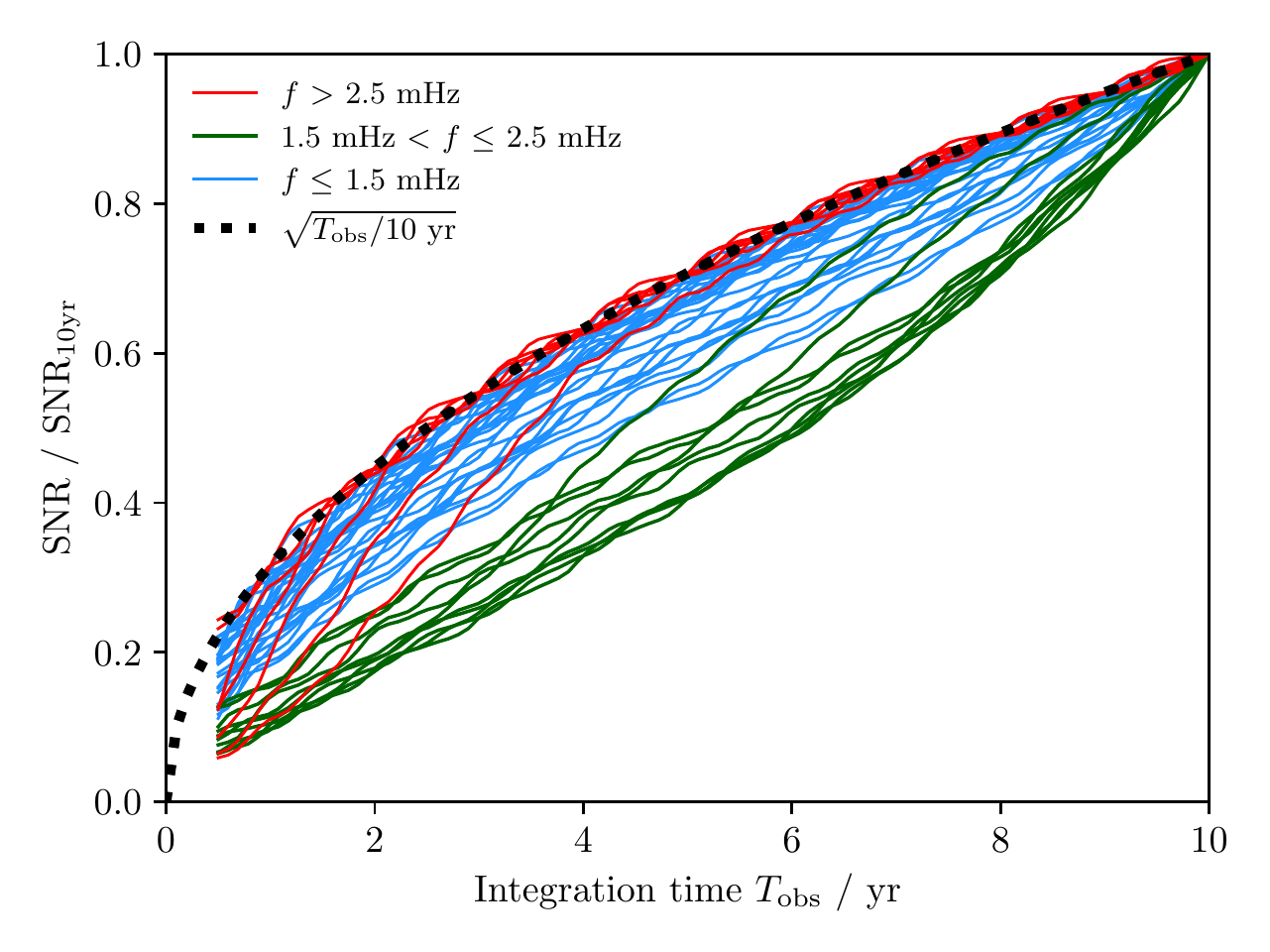}
    \caption{
    Cumulative SNR of all 43 VB candidates as a function of \LISA{} integration time over a $10\,\mathrm{yr}$ mission. 
    Sources are coloured in three categories according to their frequency. 
    The dotted black line shows the simple $\mathrm{SNR}\propto\sqrt{T_{\rm obs}}$ dependence expected for a stationary detector with a constant noise PSD. 
    This is not realised for \LISA{} because the confusion noise decreases with time; sources at intermediate frequencies where the confusion is the dominant noise source accumulated SNR approximately linearly in time.
    The oscillations visible at a frequency of $2\,\mathrm{yr}^{-1}$ are due to the orbital motion of the \LISA{} constellation.
    }
    \label{fig:SNRvsTobs}
\end{figure}

The optimal SNR over an observation time $T_{\rm obs}$ is defined as a sum over all three TDI channels
\begin{equation}
    \mathrm{SNR}_{T_{\rm obs}} = \left[\sum_\alpha\langle h|h \rangle_\alpha \right]^{1/2},
\end{equation} 
and is used to determine whether a particular VB can be detected.
The SNR increases with mission duration $T_{\rm obs}$, but not in the usual $\propto\sqrt{T_{\rm obs}}$ manner.
The rate at which the SNR accumulates is complicated by the Galactic confusion noise (see Appendix~\ref{sec:noisecurves}) that decreases over time as it becomes possible to individually resolve more of the UCB sources in the Galaxy. 
This means the effective noise PSD decreases for longer mission durations, thereby raising the SNR.
This effect is most pronounced for sources with intermediate frequencies around $\sim 2\,\mathrm{mHz}$ where the confusion is the dominant source of noise.
The rate at which the SNR accumulates is further complicated by the orbital motion of the \LISA{} constellation around the Sun. 
The quadrupolar antenna pattern of \LISA{} introduces oscillations in its sensitivity to GWs from a particular sky direction at a frequency of $2\,\mathrm{yr}^{-1}$.
These effects can be seen in Fig.~\ref{fig:SNRvsTobs} where the cumulative SNR of each candidate VB is plotted over a $10\,\mathrm{yr}$ mission duration.
This is consistent with previous work on the SNR evolution over time~\citep{Kupfer:2018jee, Seoane:2021kkk}, which also shows an SNR scaling that differs from a simple square root dependence. 
The rate at which the SNR accumulates in the first few months of the \LISA{} mission is investigated in more detail in Section~\ref{sec:LISA_early}.

The parameter estimation results from the GW-only analysis are summarised in Table~\ref{tab:GW_table}. 
\LISA{} $\mathrm{SNR}_{T_{\rm obs}}$ values are reported for mission durations of $T_\mathrm{obs} = 1$, 4, and $10\,\mathrm{yr}$ for all 43 candidate VBs.
Although parameter estimation analyses were performed for all three mission durations, results in the table are quoted for the \LISA{} nominal mission duration of $4\,\mathrm{yr}$~\citep{2017arXiv170200786A}. 
We find that UCB sources below a threshold SNR of $\lesssim 6$ generally cannot be detected or characterised by \LISA{} (the posteriors are typically broad, with no clear peaks and with an amplitude consistent with zero).
Throughout this paper, we adopt a value of $\mathrm{SNR} = 6$ as a fiducial threshold required for detection.
Parameter estimation results are only shown in Table~\ref{tab:GW_table} for the 25 resolvable sources with $\mathrm{SNR}_{4\mathrm{yr}} > 6$.

Each VB was injected into a zero-noise realisation and analysed individually.
The analysis was performed using the likelihood function in Eq.~\eqref{eq:log_like} with the noise PSD being the sum of the instrumental noise~\citep[as described in the latest \LISA{} Science Requirement Document][]{2021arXiv210801167B} and the astrophysical confusion noise from the unresolved Galactic binaries~\citep[modelled using Eq.~\ref{eq:conf_psd} from][]{Babak:2017tow}. 

Each VB is described by eight parameters: a GW strain amplitude $\mathcal{A}$, a GW frequency $f_0$ (equal to twice the binary orbital frequency, $f_0 = 2/P$), a time derivative of GW frequency $\dot{f}$, an inclination angle $\iota$, an ecliptic latitude and longitude $(b,l)$, and the initial phase $\phi_0$ and polarisation angle $\psi$. 
The injected values, where possible, are derived from the median EM-observed values in Table~\ref{tab:EM_table}. 
The GW strain amplitude is given by~\citep{Blanchet:2013haa}
\begin{equation}\label{eq:amp}
    \mathcal{A} = \frac{2(G\mathcal{M}_c)^{5/3}(\pi f_0)^{2/3}}{D c^4} ,
\end{equation}
where $\mathcal{M}_c=(m_1 m_2)^{3/5}/(m_1+m_2)^{1/5}$ is the chirp mass, and $D$ is the distance to the source.
For the injection, the frequency derivative was assumed to be driven by GW emission and has the quadrupole-formula-derived value of
\begin{equation}
    \dot{f} = \frac{96}{5} \left( \frac{G\mathcal{M}_c}{c^3} \right)^{5/3} \pi^{8/3} f_0^{11/3}.
\end{equation}
This constraint was not used in the recovery where $\dot{f}$ was treated as a free parameter, allowing us to potentially measure the effects of tides or mass transfer on the evolution of the binary.
No constraints on the initial phase and polarisation are obtained from the EM observations; the injected values for these parameters were drawn randomly from the priors.

A blind search was performed for each VB.
We emphasise that this analysis is deliberately \emph{not} taking into account what is already known about the VBs from EM observations.
The analysis used the following flat priors: $\mathcal{A} \in [0, 10\mathcal{A}_0]$; $f_0 \in [f_0 - 3/T_\mathrm{obs}, f_0 + 3/T_\mathrm{obs}]$; $\dot{f} \in [-10^{-15}, 10^{-15}]\,\mathrm{s}^{-2}$; $\cos(\iota)\in [-1, 1]$; $\sin(b) \in [-1, 1]$; $l\in [0, 2\pi]$; $\phi_0 \in [0, 2\pi]$; $\psi \in [0,\pi]$.

The modular design of the \BALROG{} code allows it to be called with a wide range of stochastic samplers. In this paper, the \textsc{nessai}~\citep{Williams:2021qyt} implementation of the nested sampling algorithm~\citep{Skilling:2006gxv} was used to sample the posterior distribution.
The runs in this section were performed with 2000 live points and required an average of 967, 2575, and 8026 CPU seconds for the 1, 4, and $10\,\mathrm{yr}$ analyses respectively.

We find that, with GW measurements only, we can generally measure the (above-threshold) VB amplitudes to the $1\%$ level, the GW frequency to a  subfrequency bin width precision (one part in $\gtrsim 10^5$), and for sources with $f_0 \gtrsim 2\,\mathrm{mHz}$ we can constrain the frequency derivative away from zero.

Compared to the recent study by \citet{Kupfer:2018jee}, our analysis includes 12 new (recently discovered) sources that satisfy the chosen resolvability criteria.
This includes five ZTF sources (one AMCVn, four DWDs), four SDSS DWDs, the DWDs SMSSJ0338 and PTFJ0533, and the UCXB 4U1820-30.
The close-to-threshold sources in \citet{Kupfer:2018jee} (PTF1919, CX1751, and CDm30) do not meet the resolvability criteria used in this study.
We see good agreement with the $4\,\mathrm{yr}$ SNRs reported in table 3 of \citet{Kupfer:2018jee}, with the exception of HMCnc (our SNR is a factor of $\sim 2$ smaller) and SDSSJ0935 (our SNR is a factor of $\sim 2$ bigger). 
These differences are due to choices for the distance to the source.
There is also broad agreement with the measurement uncertainties on the VB amplitude and inclination. 
However, in a few cases we find errors smaller than those previously reported; this is likely to be due to choices for the source distance and differences in the details of the analysis.

\begin{figure*}
    \centering
    \includegraphics[width=\textwidth]{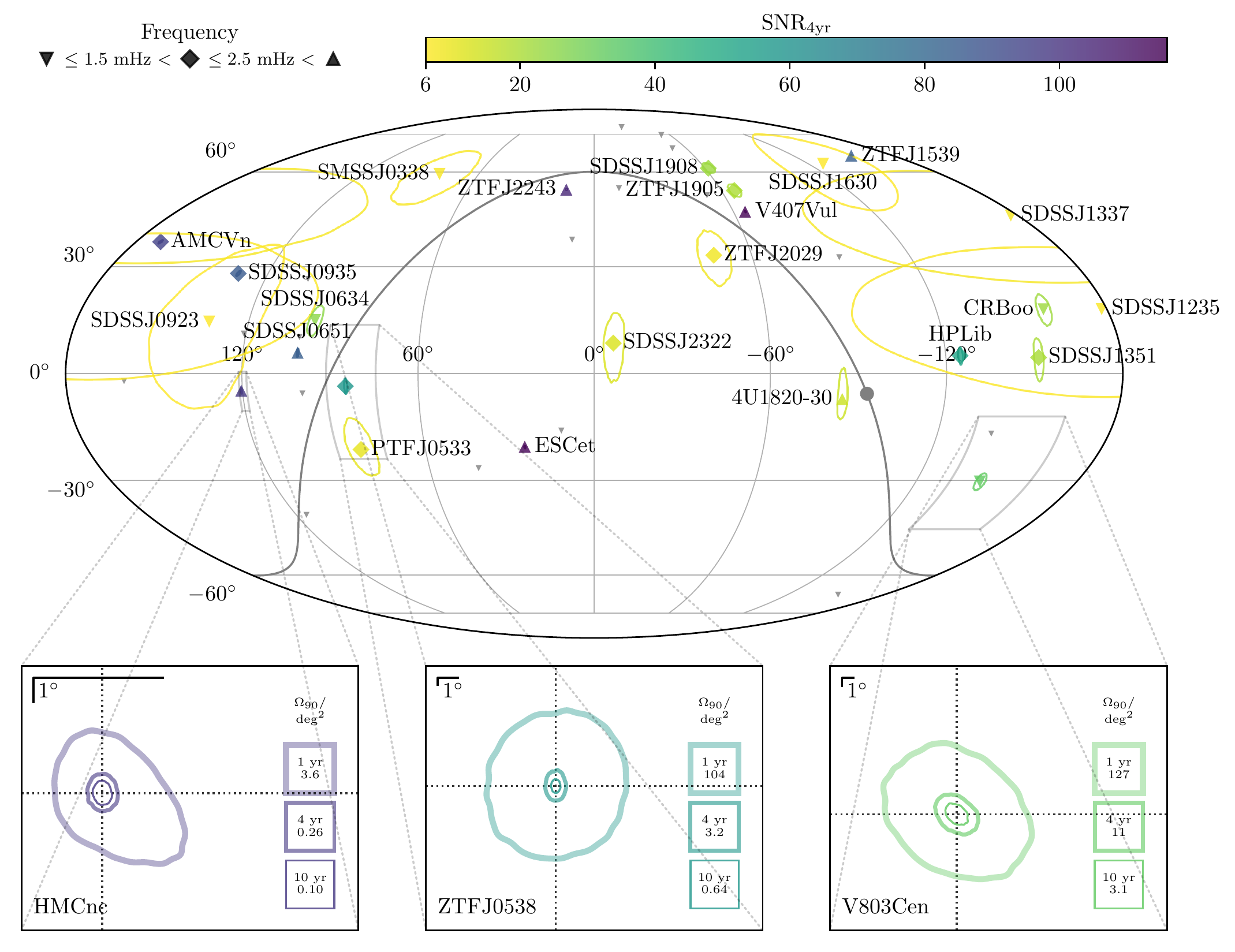}
    \caption{
    Skymaps of all 43 VB candidates in the ecliptic coordinate frame.
    Known sky positions (used as the injected values in the analysis) are indicated by markers. The marker shape indicates the VB frequency.
    For the 25 resolvable VBs satisfying $\mathrm{SNR}_{4\mathrm{yr}} > 6$, the marker colour indicates the VB $\mathrm{SNR}_{4\mathrm{yr}}$. 
    Additionally, for the above-threshold sources, the recovered sky position (90\% credible region) after $T_\mathrm{obs} = 4\,\mathrm{yr}$ is shown by a contour.
    For reference, the Galactic equator and centre are indicated by the grey line and marker, respectively.
    \textit{Insets:} For HMCnc (high $f_0$), ZTFJ0538 (medium $f_0$), and  V803Cen (low $f_0$), the recovered sky position is shown for $T_\mathrm{obs} = \{1, 4, 10\}\,\mathrm{yr}$ with the associated decreasing areas, $\Omega_{90}$, showing the improvement over time. 
    The scale of each inset is indicated in the top left by $1^{\circ}$ lines in both longitude and latitude.
    }
    \label{fig:skymap}
\end{figure*}

Fig.~\ref{fig:skymap} shows the recovered sky positions of the 25 VBs with $\mathrm{SNR}_{4\mathrm{yr}} > 6$. 
For each VB, the 90\% credible GW-recovered sky position is shown along with the injected value (which comes from the EM observations). 
The GW-derived sky position is consistent with the much more precise EM-derived sky position in all cases.
Table~\ref{tab:GW_table} also reports the 90\% credible GW-recovered sky area $\Omega_{90}$ for these sources.

\subsection{The double pulsar PSR J0737-3039}\label{subsec:double_pulsar}

The term ``verification binary'' is used for any \LISA{} source that can be observed electromagnetically in advance. 
These are mostly DWDs but, in principle, can include other compact objects. 
An example
that is tantalisingly close to being within the sensitivity reach of \LISA{} is the double pulsar PSR J0737-3039~\citep{Burgay:2003jj, Lyne:2004cj}.
The binary is mildly eccentric, $e=0.088$, so it radiates GWs at multiple frequency harmonics $f_{\rm GW}=n/P=n\times 0.113\,\mathrm{mHz}$, where $P=2.45\,\mathrm{h}$ is the orbital period (see the top panel of Fig.~\ref{fig:CharacteristicStrain}). 
The orbital parameters of this source are known with exquisite accuracy~\citep{Kramer:2021jcw} and the expected SNRs in the first few harmonics ($n=1$ to $6$) are 0.01, 1.33, 0.70, 0.20, 0.04, and 0.01 after $T_{\rm obs}=10\,\mathrm{yr}$ of \LISA{} observations.
The total SNR across all harmonics is 1.52. 
This is too quiet for \LISA{} to detect and therefore this is not expected to be a verification source.

\subsection{Verification binaries in the early months of LISA operations}\label{sec:LISA_early}

VBs will be key in helping to establish the early performance of the instrument in comparison to pre-launch predictions. This will be particularly important during the first few months of science operations.  In turn, this may well play a role in determining the timing and content of the first \LISA{} data releases.

\begin{figure}
    \centering
    \includegraphics[width=\linewidth]{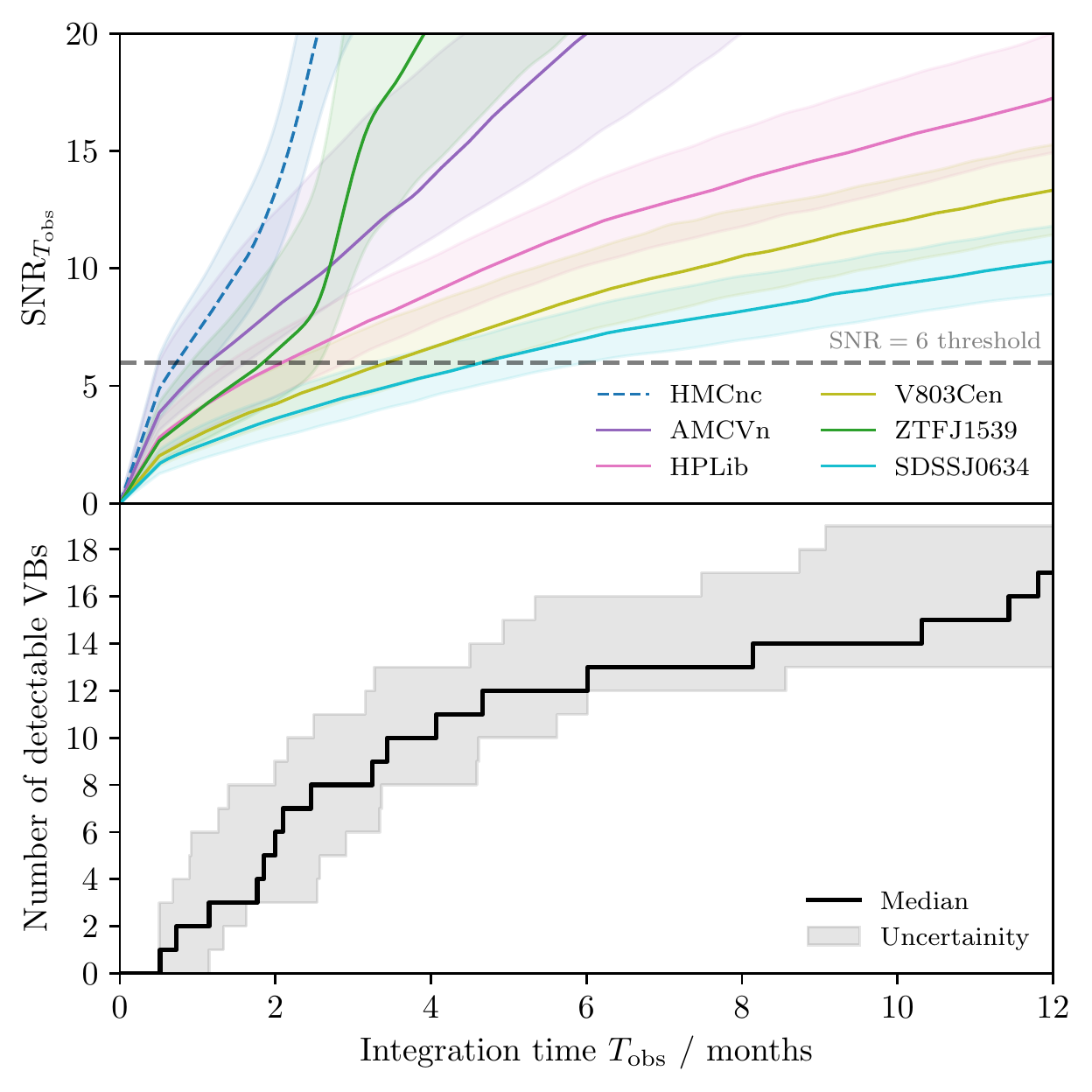}
    \caption{
    \textit{Top:} Accumulated SNR as a function of mission duration for an illustrative selection of loud VBs. The shaded region for each VB corresponds to the 50\% confidence interval, with the uncertainty coming from the astrophysical parameters and \LISA{} orbit. 
    We note HMCnc (shown here with a dashed line) has a particularly large uncertainty on its distance; we account for this uncertainty by drawing samples in the range $5 \text{--} 10\,\mathrm{kpc}$. 
    Even with this accounted for, its high frequency means that it is likely to be one of the loudest VBs (which is why we include it here).
    \textit{Bottom:} Number of VB sources that exceed the detection threshold $\mathrm{SNR}>6$ as a function of observation time. 
    The shaded region in the bottom panel corresponds to the same 50\% confidence interval from the top panel (coming from both the astrophysical and orbital uncertainties). 
    None of the VBs below $1\,\mathrm{mHz}$ are detectable within the first year of \LISA{} operation, as seen in the bottom panel of Fig.~\ref{fig:CharacteristicStrain}.
    }
    \label{fig:SNR_uncertainity}
\end{figure}

To help guide expectations for which and how many VBs might be detectable in the early months of \LISA{} science operations, in the top panel of Fig.~\ref{fig:SNR_uncertainity} we show the SNR evolution over the first year of the mission for an illustrative selection of loud VBs.  
The total number of VB sources that exceed the $\mathrm{SNR} > 6$ detection threshold as a function of mission duration is plotted in the bottom panel.

For the SNR calculations, the GW signals are computed with astrophysical parameters derived from Table \ref{tab:EM_table}. 
Two sources of uncertainty are accounted for in these $\mathrm{SNR}$ calculations: astrophysical and orbital. 
The astrophysical uncertainties arise from the errors on the parameters obtained from EM observations.
The orbital uncertainty reflects the fact that we do not yet know the exact positions of the \LISA{} spacecraft at the start of science operations. 
Over multi-year observations, the size of the orbital uncertainty decreases as the motion of the \LISA{} constellation averages over a complete orbit; however, for observation times shorter than a year this is an important extra source of uncertainty.

Fig.~\ref{fig:SNR_uncertainity} accounts for both sources of uncertainty using a Monte Carlo average. The SNR was computed as a function of $T_{\rm obs}$ for a set of $1000$ parameter draws (astrophysical and orbital position) for each VB. The astrophysical parameters were drawn either from a Gaussian distribution (for parameters where both the mean and standard deviation are available in Table~\ref{tab:EM_table}) or from a uniform distribution in cases where parameters are unconstrained (to be conservative, for sources with inclination given as [60] we draw inclination samples uniformly in the range $0 \text{--} \pi$). 
The initial orbital position of \LISA{} is described by two angles~\citep[see, for example, the appendix of][]{Cornish:2002rt}: one describing the phase of the centre of the constellation around the Sun and one describing the orientation of the three spacecraft within the constellation. 
These angles were drawn uniformly across their full ranges. 

The size of the orbital SNR uncertainty is larger than the astrophysical SNR uncertainty at early times, but is smaller for mission durations $T_{\rm obs}\gtrsim 3\,\mathrm{months}$.
These SNR calculations are assuming that the instrumental noise requirements for the \LISA{} mission~\citep{2021arXiv210801167B} are met exactly, and that the Galactic confusion noise from the unresolved Galactic binaries is described by Eq.~\eqref{eq:conf_psd} from \citet{Babak:2017tow}; in reality, both of these are additional sources of SNR uncertainty.
    
From Fig.~\ref{fig:SNR_uncertainity}, it is expected that two VBs will likely be observable after just 1 month of observations. After 6 (12) months, it is expected that at least 11 (13) VBs will be detectable.
Further results for the time to detection for all 25 individual VBs are shown in the bottom panel of Fig.~\ref{fig:CharacteristicStrain}, from which one can see that none of the VBs below $1\,\mathrm{mHz}$ are expected to be observed before \LISA{} gathers $\approx 2.5\,\mathrm{yr}$ of data.

\section{Electromagnetic -- gravitational-wave Synergies}\label{sec:EMpriors}

The analysis in the previous section deliberately did not use any prior EM-derived knowledge of the VB parameters.
This information can be included in the prior of the GW analysis where it may help to confidently detect a VB close to the SNR threshold or improve the parameter estimation accuracy of a louder VB.
This section explores the influence of prior knowledge of the frequency ($f_0$), inclination ($\iota$), and sky location ($l,b$). 
The choice of these parameters was motivated by the availability of EM measurements and on expectations from previous (Fisher-matrix) studies~\citep{Shah:2012vc, Shah:2013ema}.
The analysis was originally performed on a set of simulations where $f_0$, $\iota$, $l$, and $b$ were separately fixed to the respective EM-measured value. 
We then analysed the simulations where combinations of these variables were fixed. 
In order to have clearer results, only the most informative constraints are presented in this section.
We illustrate our results with two example VBs: SDSSJ0651 that is nearly edge-on ($\iota=87\,\mathrm{deg}$; this gives rise to approximately linearly polarised GWs), and SDSSJ1908 which is nearly face-on ($\iota=15\,\mathrm{deg}$; this gives rise to approximately circularly polarised GWs).

For a specific mission duration, five different types of analysis were considered: (1) a blind search (using the uninformative priors from Section~\ref{sec:VBstudy}) and four searches with various parameters fixed to their EM-measured values. These are: (2) frequency fixed, (3) inclination fixed, (4) sky position fixed, and (5) sky position and inclination fixed.
Search (4) can be described as a `directed' or `targeted' search.

The effect of a particular prior choice is illustrated by considering its effect on the one-dimensional marginalised posterior on the amplitude, $\mathcal{A}$.
We use the ratio of the mean, $\mu$, and the standard deviation, $\sigma$, of this amplitude posterior as a proxy for how confidently a VB source can be detected.
At low SNR, the amplitude is consistent with zero and resembles a truncated distribution; this has a ratio $\mu/\sigma\approx 1$, meaning that the VB cannot be detected.
At high SNR, the amplitude posterior is peaked away from zero and has a ratio $\mu/\sigma\gg 1$, meaning that the VB can be detected.
Empirically, we identify a threshold value of $\mu/\sigma > 2.5$ as being the minimum value necessary to detect a VB source (this corresponds roughly to an SNR of 6).

\begin{figure}
    \centering  
    \includegraphics[width=\linewidth]{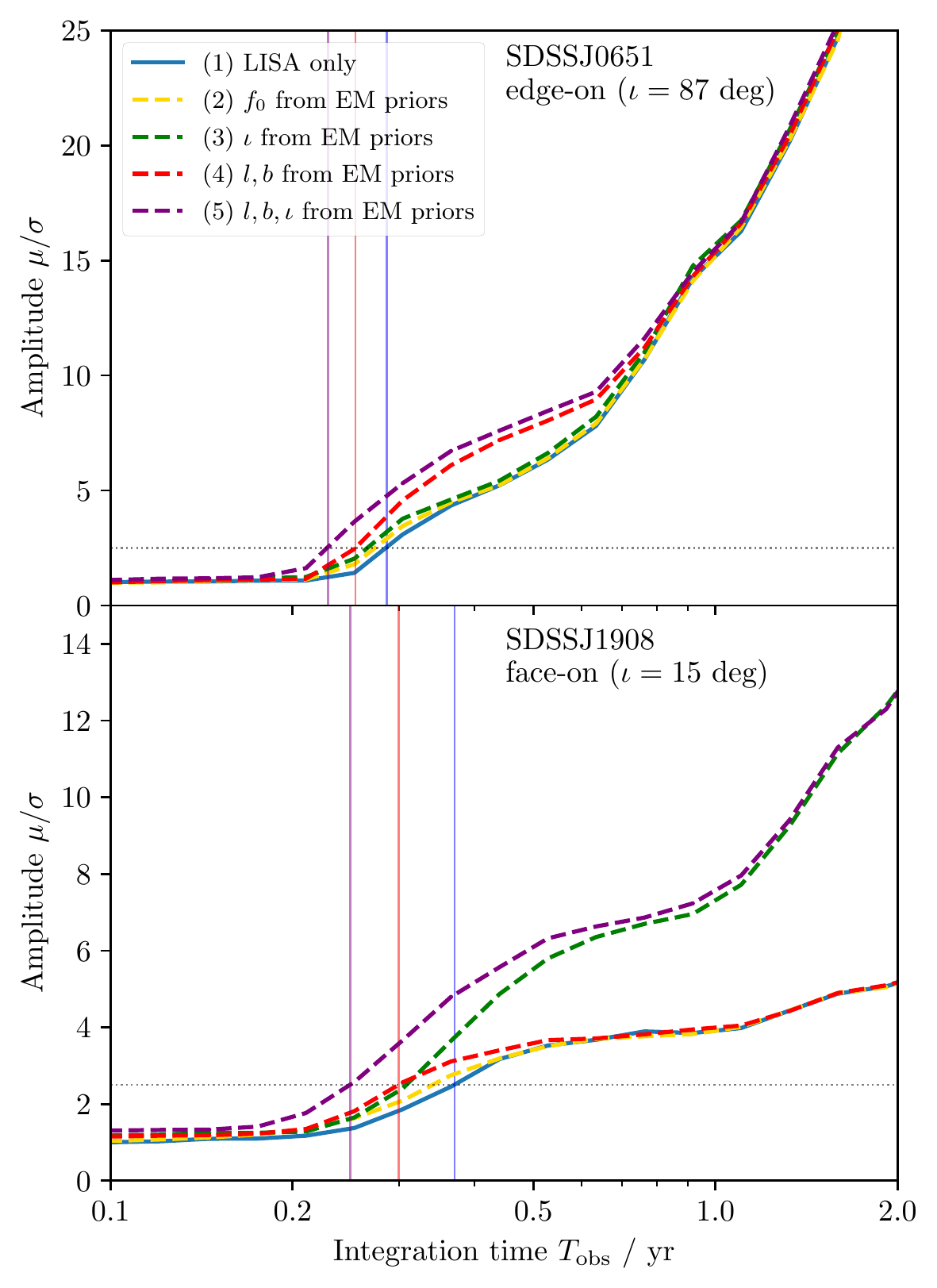}
    \caption{
    Mean over standard deviation of the amplitude posteriors for the edge-on source SDSSJ0651 (top), and for the face-on source SDSSJ1908 (bottom) as a function of mission duration. 
    The solid blue line shows the results for a blind search (1), while dashed lines show searches with different EM-informed priors $(2 \text{--} 5)$. 
    The horizontal line indicates the detection threshold $\mu/\sigma=2.5$. 
    Vertical lines show the detection times for the (1), (4), and (5) searches. 
    In both cases, prior knowledge of VB parameters reduces the time to detection. 
    For long $T_{\rm obs}$, prior knowledge of the inclination for the face-on source leads to a very significant improvement in the posterior.
    }
    \label{fig:MeanStd}
\end{figure}

Fig.~\ref{fig:MeanStd} shows the evolution of the ratio $\mu/\sigma$ with increasing mission duration for the two example sources. 
The earliest time at which the source can be detected is when $\mu/\sigma = 2.5$.
Compared to the blind search (1), in all other cases $(2 \text{--} 5)$ the effect of including EM prior information is a modest reduction in the time to detection (knowledge of the inclination parameter leads to the biggest reduction).
SDSSJ0651 (SDSSJ1908) is detected after $104\,\mathrm{d}$ ($134\,\mathrm{d}$) with a blind search, $92\,\mathrm{d}$ ($109\,\mathrm{d}$) with the directed search, and $83\,\mathrm{d}$ ($91\,\mathrm{d}$) with search (5).

From a Bayesian perspective, the reduction in time to detection with improved prior knowledge is expected and can be explained in terms of the Bayes' factor and Occam's razor. Fixing a model parameter to its true value reduces the size of the remaining parameter space, increasing the Bayesian evidence for the signal hypothesis by reducing its associated Occam penalty. The evidence for the null hypothesis is unaffected, so the Bayes' factor (equal to the ratio of the evidences) increases, making detection easier.

After the source is detected, the two panels of Fig.~\ref{fig:MeanStd} show different behaviour. For the edge-on source, prior knowledge from EM observations does not significantly affect the amplitude measurement. However, for face-on sources prior knowledge of the inclination angle $\iota$ does lead to a dramatic improvement in the amplitude measurement; this improves further for longer observations.
For SDSSJ1908, after $T_{\rm obs}=2\,\mathrm{yr}$, prior knowledge of $\iota$ leads to an improvement in the amplitude measurement by a factor of 2.4.
Similar results were obtained for all the other VBs that were analysed.

From the above discussion, it is clear that prior EM-derived knowledge of the inclination angle is particularly important.
The reason for this can be traced back to the fact that two-dimensional posteriors on $\mathcal{A}$ and $\iota$ typically show a strong degeneracy; this is particularly true for nearly face-on  or face-off sources.
Therefore, fixing the inclination to its true value helps to improve the measurement of the amplitude. 
The amplitude in turn is related to the component masses and the distance to the source (see Eq.~\eqref{eq:amp}).
This is consistent with what was found by \citet{Shah:2012vc} using Fisher matrices.

It should also be noted that the GW measurements can also be used to improve the EM measurements of the inclination angle. Even in cases where the inclination is known from EM observations (see Table~\ref{tab:EM_table}), it is typically not known in which direction on the sky the VB is orbiting (i.e.\ an EM-measured inclination of $\iota=1^\circ$ could correspond to a nearly face-on source seen orbiting in a counterclockwise direction or to a nearly face-off source with $\iota = 179^\circ$ seen orbiting clockwise). 
GW measurements will break this degeneracy (see results in Table~\ref{tab:GW_table}).
This is consistent with what was found by~\cite{Littenberg:2019mob} in the specific case of ZTF J1539+5027.

\begin{figure*}
    \centering
    \includegraphics[width=\textwidth]{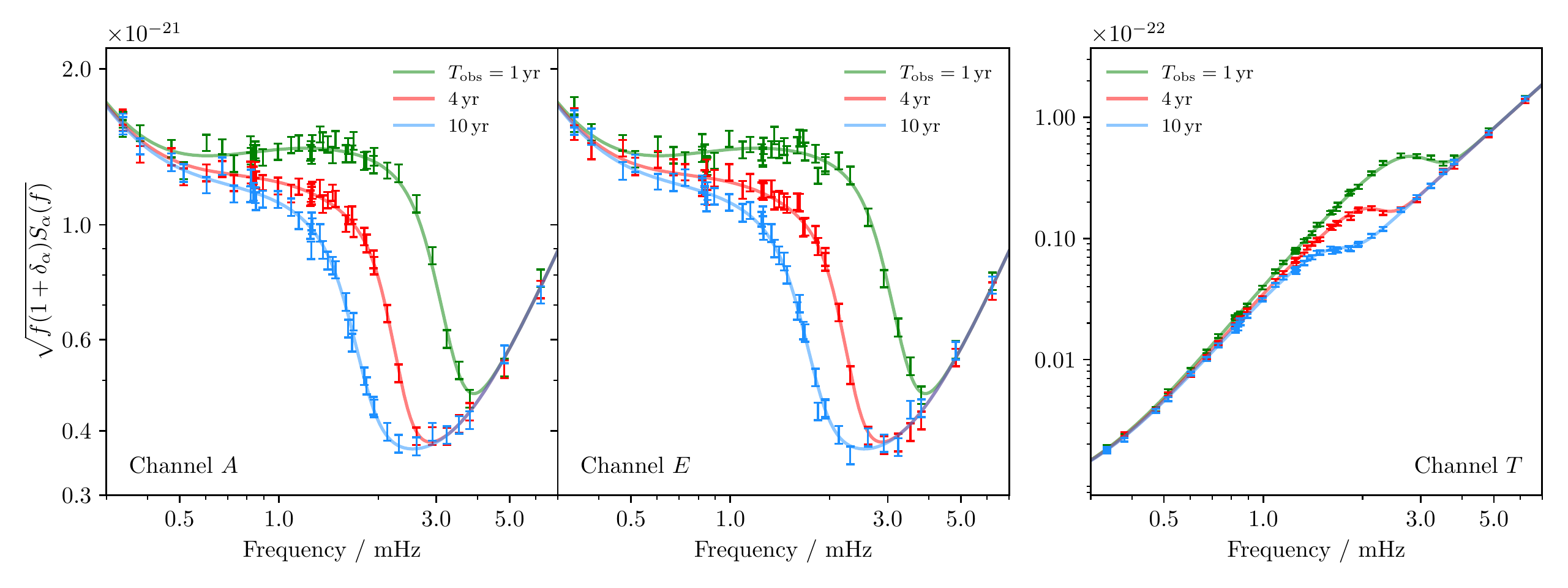}
    \caption{
    Representative constraints on the noise curve in channels $A$, $E$, and $T$, with mission durations of 1, 4, and $10\,\mathrm{yr}$ plotted in blue, orange, and green respectively. 
    Error bars show 90\% confidence intervals.
    Also plotted for comparison is the \LISA{} noise PSD used for the injection.
    At low frequencies the GW signal in the $T$ channel is highly suppressed.
    }
    \label{fig:NoiseBoxPlots}
\end{figure*}

Knowledge of the inclination helps most for face-on systems. However, it is harder to measure $\iota$ for such systems because they are not eclipsing. 
If the inclination is not known but the VB source is known \emph{not} to be eclipsing, this implies that the binary is not close to edge-on and therefore places a weak constraint on the inclination.
To investigate the usefulness of such a constraint we conducted a post-analysis on all VBs with $\iota=[60^\circ]$ in Table~\ref{tab:EM_table}. Posterior samples with $85^\circ<\iota<95^\circ$ were removed, mimicking the effect of a prior that excludes this range of $\iota$ that would give rise to eclipses. Unfortunately, the resulting posteriors showed a negligible improvement.

\section{Accounting for unknown noise levels}\label{sec:unknown_noise}

The VB analyses described above all used the standard form of the likelihood, Eq.~\eqref{eq:log_like}.
This assumes that the noise in each independent data time series is additive, stationary, and Gaussian. 
The statistical properties of this type of noise can be described in terms of the PSD, $S_\alpha(f)$.
The standard GW likelihood also assumes that this noise PSD is known exactly beforehand.
In reality, the instrumental noise sources will not be understood perfectly and it will also be necessary to account for the presence of the Galactic foreground as an additional uncertain noise source in the analysis.
In this section we describe how this can be done as part of a fully Bayesian analysis of VB sources by relaxing the assumption of a known noise PSD.

Hereafter, instead of Eq.~\eqref{eq:log_like}, we use a modified GW likelihood:
\begin{align} \label{eq:log_like_unknownS}
    P(d | h) = \prod_\alpha \frac{\exp\left(-2\sum_k \frac{|d_\alpha(f_k) - h_\alpha(f_k)|^2}{(1+\delta_\alpha)S_\alpha(f_k)}\delta f\right)}{2\pi\prod_k (1+\delta_\alpha)S_\alpha(f_k)\delta f}\,,
\end{align}
where $\alpha$ labels the TDI channel.
When the noise is treated as an unknown in the analysis it is necessary to include the denominator of Eq.~\eqref{eq:log_like_unknownS}~\citep[see, e.g.,][]{Littenberg:2011zg}.

A fixed, reference PSD $S_\alpha(f)$ is used in Eq.~\eqref{eq:log_like_unknownS} [although we use the same symbol, we stress that the meaning of $S_\alpha(f)$ here is different from that in Eq.~\eqref{eq:inner_prod}]. 
This can be chosen to be our best \emph{a priori} estimate for the \LISA{} noise; this was taken to be the same PSD that was used in Section~\ref{subsec:parameter_estimation}, including the estimate of the confusion noise.
This is fixed throughout the analysis.
However, this reference noise PSD is multiplied by a factor $1+\delta_\alpha$.
The three new parameters $\delta_\alpha$ describe variations in the noise level relative to the \emph{a priori} PSD estimate. These parameters can account for both instrumental and Galactic foreground noise sources.
VBs are extremely narrow band sources with GW signal power present only in a few frequency bins; therefore, we choose to use a single parameter in each TDI channel to describe variations in the noise level only (the shape of the PSD is kept fixed).
Flat priors were used on the $\delta_\alpha$ parameters with sufficiently broad ranges that the posteriors are unaffected by prior boundaries.

In Section~\ref{sec:confusion} this likelihood is used to analyse multiple signals simultaneously. 
First, however, the VB sources were reanalysed individually with this likelihood to ensure they can still be individually detected and characterised in the presence of unknown noise levels.
The VBs were injected into simulated \LISA{} noise generated from the PSDs shown in Fig.~\ref{fig:NoiseBoxPlots} (see also the solid black curve in the top panel of Fig.~\ref{fig:CharacteristicStrain}).

Our ability to constrain the PSD $\delta_\alpha$ parameters depends on how much data are analysed.
The sums and products over $k$ in Eq.~\eqref{eq:log_like_unknownS} are taken over a small frequency range centred on the GW frequency of the VB.
Because $\delta_\alpha$ describes the noise level across the whole band, the more frequency bins that are included the better the resulting measurement of $\delta_\alpha$. 
In order to ensure a fair comparison between $\delta_\alpha$ posteriors from different analyses the number of frequency bins was fixed to 438 in all cases. 
This value was chosen to be as small as possible while still comfortably containing all the signal power for all the VB sources (accounting for Doppler broadening and a $10\,\mathrm{yr}$ frequency drift for those sources with large $\dot{f}$).
All other details of this analysis are the same as those presented in Section~\ref{sec:VBstudy}.

For each of the 43 VB candidates, two analyses were performed at mission durations of $T_{\rm obs}=1$, 4, and $10\,\mathrm{yr}$: an analysis with known PSDs (all three $\delta_\alpha$ parameters fixed to zero) and an analysis with unknown noise levels ($\delta_\alpha$ allowed to vary).
Notice that even in cases where the VB cannot be detected, it is still possible to measure $\delta_\alpha$.
For those VBs above the threshold, we find that it is still possible to detect and characterise them in the presence of unknown noise levels. The recovered posteriors were nearly identical in both simulations for most VBs, with a small number showing a small shift in some parameters, consistent with the width of the posterior. 

Two interesting cases were CX1751 after $4\,\mathrm{yr}$ and ZTFJ0640 after $10\,\mathrm{yr}$. 
Here the SNRs were just below threshold, with SNRs of 4.89 and 4.34 respectively.
This resulted in VB parameter posteriors that were somewhat constrained when the $\delta_\alpha$ were fixed but that became unconstrained when the $\delta_\alpha$ were free parameters; posterior information for marginally subthreshold sources can be lost when marginalising over uncertain noise levels. 

It is also possible to use these results to gain some insight into our ability to measure the \LISA{} noise level.
Because the $\delta_\alpha$ are treated as free parameters we obtain posterior distributions on these parameters for all 43 VB candidates.
Plotted in Fig.~\ref{fig:NoiseBoxPlots} are the 90\% confidence regions for the reconstructed noise level $\sqrt{f(1+\delta_{\alpha})S_\alpha(f)}$ in all three TDI channels.
The results are shown for all 43 separate VB candidate analyses on a single plot as a function of frequency. 
For each VB (and for each $T_{\rm obs}$) we have single measurement of $\delta_\alpha$.
We plot the posterior on the noise curve $\sqrt{f (1+\delta_\alpha) S_\alpha(f)}$.
Because we have VBs spread across a range of frequencies, this set of measurements can be used as a crude reconstruction of the full noise PSD across the whole \LISA{} band.
Because we are analysing VBs one at a time (with no other sources present) and in the presence of simulated Gaussian noise, we expect that the recovered values of $\delta_\alpha$ should be consistent with the PSD used for the injection; this can be seen to be the case in Fig.~\ref{fig:NoiseBoxPlots}. 
With our choice of number of frequency bins, $N_\mathrm{bin} = 438$, the noise curve can be measured to an accuracy of $\approx 8\%$ (90\% credible interval).
We stress that the magnitude of the uncertainties on the PSDs shown in Fig.~\ref{fig:NoiseBoxPlots} are determined by our choice of $N_\mathrm{bin}$ and the uncertainty on $\delta_\alpha$ scales as $1/\sqrt{N_\mathrm{bin}}$. We also stress that we have taken the most conservative approach in which the three noise parameters are treated as independent. 
Understanding of \LISA{}'s subsystem behaviour during mission operation may provide additional constraints across parameters describing the noise.

\section{Accounting for source confusion}\label{sec:confusion}

Thus far, the VBs have been treated in isolation.
This neglects the rest of the GP of UCBs that \LISA{} will face, both individually resolved and unresolved. 
(However, the effects of the unresolved sources are partly accounted for in the model for the Galactic confusion noise.)
In this section, we adopt a fiducial mock Galaxy catalogue to directly determine the impact of the rest of the UCBs in the Galaxy on our ability to measure VB parameters. 

Here we consider the GP of detached DWDs only, as they are expected to be at least an order of magnitude more numerous than the other types of stellar remnant binaries in \LISA{}'s frequency band~\citep{LISA:2022yao}. Specifically, we use a mock DWD catalogue from \citet{Wilhelm:2020qjc} assembled by combining the DWD binary population synthesis model of \citet{Toonen:2012jj} with a snapshot of {\sc GALAKOS}, a high-resolution $N$-body simulation of a stellar disc and bulge/bar with structural parameters that reproduce the currently observed properties of our Galaxy~\citep{DOnghia:2020}. Representative of a Milky Way-like Galaxy with a total stellar mass of $5\times 10^{10}\,$M$_\odot$, the catalogue contains $\sim 1.4\times10^7$ DWDs emitting in the \LISA{} frequency band. Based on an SNR criterion, \citet{Wilhelm:2020qjc} showed that $\sim 2.2\times10^4$ DWDs could be detected within $4\,\mathrm{yr}$ of the mission, while the rest would contribute to the Galactic confusion foreground signal.

\subsection{Population analysis}\label{subsec:pop_analysis}

First, we attempt to quantify how `confused' each VB source is.
This is designed to be a measure of both how many other UCBs are close in frequency to the VBs, and how loud these sources are.
A source from the GP is classified as `close' if its frequency $f_0^\mathrm{GP}$ (accounting for Doppler broadening) crosses into any of the 10 frequency bins (for a $4\,\mathrm{yr}$ mission) centred on the VB initial frequency $f_0$; i.e.\ if
\begin{equation} \label{eq:filt_condition}
    \left|f^\mathrm{GP}_0 - f_0\right| < 5\delta f + \frac{v}{c}f_0 ,
\end{equation}
where $v/c\approx 10^{-4}$ is the orbital velocity of \LISA{} around the Sun.
The actual Doppler broadening will depend on the ecliptic latitude of the source; Eq.~\eqref{eq:filt_condition} uses the maximum value.
This can be thought of as counting sources from the GP which have power in the 10 closest frequency bins to the VB.

\begin{figure}
    \centering
    \includegraphics[width=\linewidth]{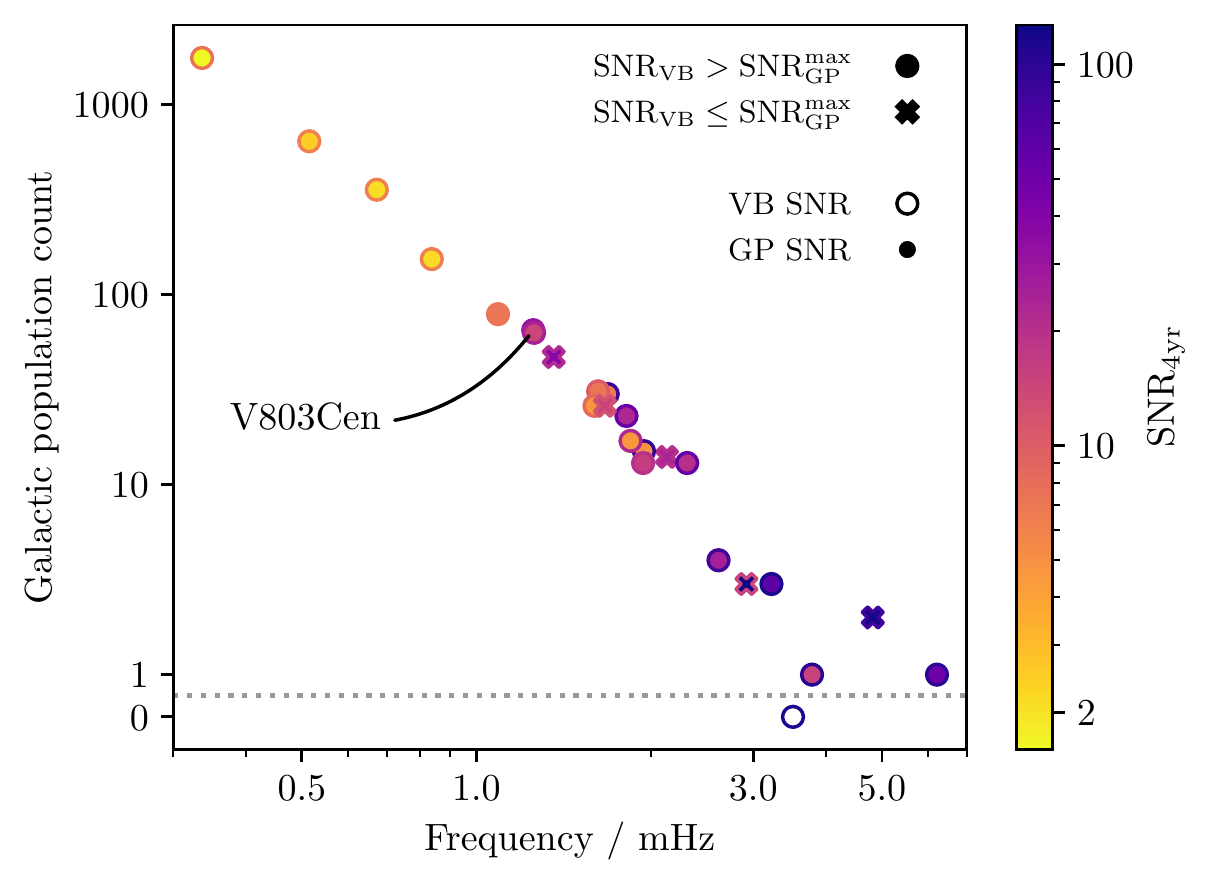}
    \caption{
    The number of DWDs in a simulated GP~\citep{Wilhelm:2020qjc} within 10 frequency bins ($T_\mathrm{obs} = 4\,\mathrm{yr}$) of each VB, after accounting for Doppler broadening (see Eq.~\ref{eq:filt_condition}). 
    These sources will need to be accounted for in a realistic analysis of the VB.
    Marker fill colour indicates the $\mathrm{SNR}_{\mathrm{4yr}}$ of the loudest member of the GP within the frequency range, while the marker edge colour indicates the $\mathrm{SNR}_{\mathrm{4yr}}$ of the VB. 
    We consider a VB to be more `confused' (i.e.\ likely to be harder to separate from the rest of the population) if it has more, or louder, DWD sources nearby in frequency.
    VBs with a circular marker satisfy $\mathrm{SNR}_{\mathrm{VB}} > \mathrm{SNR}_{\mathrm{GP}}^{\mathrm{max}}$, while those with a cross satisfy $\mathrm{SNR}_{\mathrm{VB}} \leq \mathrm{SNR}_{\mathrm{GP}}^{\mathrm{max}}$. 
    The VB V803Cen is highlighted; this is analysed further in Section~\ref{subsec:multisource_search}.
    }
    \label{fig:gp_snr}
\end{figure}

For each VB, members of the GP within the permitted frequency range were counted and their SNRs computed. 
Note that these SNRs are computed in the same manner as those in Table~\ref{tab:GW_table}, with respect to the fixed instrumental plus confusion noise curve.
Then, the highest SNR among the sources in the GP was identified and compared to the SNR of the VB. 
This information is summarised in Fig.~\ref{fig:gp_snr}.

As expected, the GP source count is a steeply decreasing function of frequency. 
In this particular realisation of the Galactic DWD population, the VBs CRBoo, SDSSJ2322, SDSSJ1351, 4U1820-30, and ZTFJ1539 have associated with them at least one DWD with an SNR greater than their own, i.e.\ $\mathrm{SNR}_{\mathrm{VB}} > \mathrm{SNR}_{\mathrm{GP}}^{\mathrm{max}}$.
The case of 4U1820-30 is particularly extreme with the ratio $\mathrm{SNR}_\mathrm{VB}/\mathrm{SNR}_\mathrm{GP}^\mathrm{max}=0.12$.

All VBs will be confused to some extent, with the possible exception of a few of the highest frequency VBs.
In some cases there are thousands of other sources nearby in frequency, including several that are louder than the VB itself.
In order to successfully perform a GW analysis of the VB under realistic conditions, it is therefore necessary to account for the presence of other individually resolvable DWDs from the GP (whose number is unknown \textit{a priori}) along with the VB and to marginalise over uncertain levels of the noise in the three TDI channels (to which the quiet unresolved DWDs contribute).
Optionally, it is also possible to account for the fact that the VB's sky location is known, i.e.\ to perform a ``directed search'' for the VB, where the sky position angles $l$ and $b$ are fixed to their known values (given in Table~\ref{tab:EM_table}). 
This is the goal of the following section.

\subsection{VB inference with a realistic confusion foreground}\label{subsec:multisource_search}

In this section we take as an example VB the binary V803Cen, and place it in the simulated GP described above.
V803Cen has 65 DWDs from the GP nearby in frequency (assessed using the criteria in Eq.~\ref{eq:filt_condition}), two of which have $\mathrm{SNR}_\mathrm{4yr} > 6$ (the next loudest has an SNR of 5.01).
Therefore, it might be expected that we would need to model these two additional sources to perform reliable inference on the VB.

To demonstrate this, a $4\,\mathrm{yr}$ \LISA{} data instance was generated including an instrumental noise realisation, the VB V803Cen, plus 229 other DWDs from the (mock) Galaxy. 
This number includes the 65 sources closest in frequency to the VB, plus the additional sources (in the $\sim 100$ frequency bins outside the initial frequency range) to ensure we lose no power at the edge of our frequency band used in the analysis.

\begin{figure*}
    \centering
    \includegraphics[width=\textwidth]{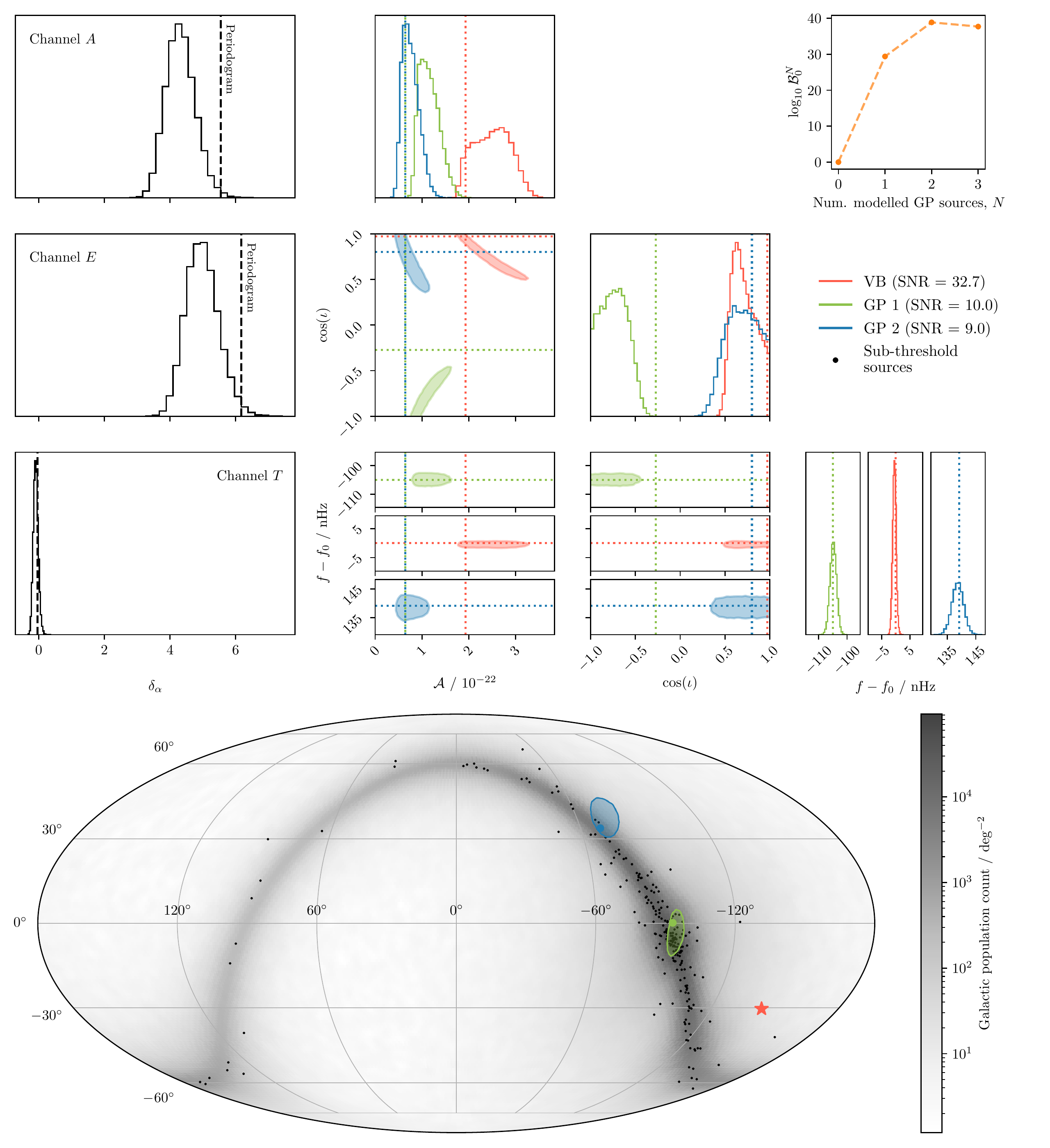}
    \caption{
    \textit{Top:} Corner plot showing posteriors for V803Cen and two (nearby in frequency) sources from the GP with $\mathrm{SNR}_\mathrm{4yr} > 6$, and posteriors on the noise level in each TDI channel with respect to the (design) instrumental noise curve (parameterised through $\delta_\alpha$). A directed search for the VB was performed (shown in red), and blind searches with no prior information for the two Galactic binaries were performed (shown in blue and green). \textit{Bottom:} The skymap of the VB V803Cen (red star) and its associated above-threshold Galactic binaries (blue and green markers), among its subthreshold Galactic binaries (black markers) and the remainder of the mock Galaxy realisation (heatmap). Contours show the recovered sky position (90\% credible region) at $T_\mathrm{obs} = 4\,\mathrm{yr}$ for the two above-threshold Galactic binaries.
    }
    \label{fig:MultiCorner}
\end{figure*}

Using this data instance, we then perform an unknown-noise parameter estimation (using Eq.~\ref{eq:log_like_unknownS}) of the VB while simultaneously modelling $N$ extra sources.
The cases $N = [0,1,2,3]$ were explored, and in all cases the VB inference is ``targeted'' (that is, with fixed sky location) while the other DWDs are searched for over the full sky. 
The total number of unknown parameters in the analysis is therefore $3+6+(8\times N)$.
All other priors are the same as those described in Section~\ref{subsec:parameter_estimation}, with the exception of $f_0$ which was set to be within the range given by Eq.~\eqref{eq:filt_condition}.
To deal with the label-switching problem, which arises when multiple sources described by the same model are included in the analysis, we follow the approach in \citet{Buscicchio:2019rir} and order sources by their frequency.
All analyses were performed with 4000 live points, and took $\sim 1$, 30, 200 and 500 CPU hours for the $N=0$, 1, 2, and 3 runs, respectively.

The results are summarised in Fig.~\ref{fig:MultiCorner}.
In the top right panel of the figure we show the log-Bayes' factors, $\log_{10}\mathcal{B}^N_0$, comparing the model evidences for different values of $N$ (normalised to the $N=0$ analysis, i.e.\ the VB-only analysis).
This peaks at $N=2$ extra sources, consistent with the expectation that only DWDs above a threshold SNR of $\sim 6$ can be detected.
However, the $N=3$ analysis also has comparable support to the $N=2$ case (it is disfavoured by $\log_{10}\mathcal{B}^2_3 = 1.2$); we speculate that this is related to the presence of another marginally subthreshold source.

The corner plot in the top half of the figure, with the coloured histograms, shows posteriors on the amplitude, inclination, and frequency for each of the modelled sources in the $N=2$ analysis.
Note that, because of the narrow frequency posteriors, the frequency panels have been split to zoom-in on the posterior for each source.
The true (injected) values for each source are shown with the dotted lines.
We see posteriors consistent with the injected values for the VB, indicating that we have successfully accounted for the unknown noise and confusion sources.
However, only one of the GP sources has posteriors consistent with the injected values.
The bias seen in the recovery of GP 1 is a result of our imperfect signal model; the inference on this source is confused by the presence of many other sources (some of which are just below the SNR threshold of 6, and so have not been completely captured by our noise model).
Comparing the results to those in Section~\ref{sec:VBstudy}, we see comparable errors in the frequency, amplitude and inclination ($\Delta f/\delta f \sim 0.08$, or one part in $2 \times 10^6$, $\Delta\mathcal{A}/\mathcal{A} \sim 0.15$, $\Delta \iota \sim 13^\circ$).

To the left of the corner plot are the posteriors on the unknown noise parameters for each TDI channel, $\delta_\alpha$, as described in Section~\ref{sec:unknown_noise}.
These represent a modification to the instrumental noise curve (no analytical description of the confusion noise is present in this analysis).
In the $A$ and $E$ channels, we measure the power from the confusion noise to be $\sim 5\pm 1$ times greater than the instrumental noise only.
The dashed line plotted on each of the histograms indicates an expected value for $\delta_\alpha$, from a periodogram-type calculation.
To do this, we generate a new data instance with an instrumental noise realisation and a population of confusion sources (in the same way the data we analysed were created), but we exclude the three modelled sources from the data.
In other words, we subtract perfectly the sources we model from the data.
With this new data instance, which represents our noise $n^\alpha(f)$, we can calculate the PSD $S_\alpha$ via
\begin{equation}\label{eq:periodogram}
    S_\alpha = \frac{2}{T_\mathrm{obs}} \left\langle |n^\alpha(f)|^2 \right\rangle.
\end{equation}
This approximates the PSD as being constant over the frequency range of the data; as our analysis data are narrow ($\sim 150$ frequency bins, with $T_\mathrm{obs} = 4\,\mathrm{yr}$), this is a reasonable approximation.
Finally, to obtain the predicted $\delta_\alpha$, we compare this measured PSD to the instrumental noise curve at the VB frequency $f_0$.
Our posteriors are broadly consistent with the prediction, but our measurement tends to lower values. 
One reason for this is that we cannot perfectly model the above threshold sources in our Bayesian analysis. 
As seen in the other panels of Fig.~\ref{fig:MultiCorner}, confusion between the GP sources leads to biases in the posteriors. 
As we may be ``absorbing'' power from other (just below threshold) GP sources into our model, this causes an underestimate of the noise. 
The loudest below-threshold sources may also break the assumption of a constant (in time and frequency) PSD in the application of Eq.~\eqref{eq:periodogram}.

Finally, the bottom panel of the figure shows a skymap with the locations of the VB V803Cen (a targeted sky search, red star), the other two DWDs we model (for which the 90\% credible regions are shown with green and blue contours), and the other 227 subthreshold sources from the GP that were included in the data.
The heatmap indicates the density of all $\sim 1.4 \times 10^7$ sources in the (mock) Galaxy. 
We see sky recovery consistent with the injected values.

\section{Discussion and Conclusions}\label{sec:conclusion}

Unique among GW sources, a small number of Galactic UCBs are known in advance of the operation of a GW observatory as guaranteed sources.
Extending previous studies of VBs, e.g.\ \citet{Kupfer:2018jee}, we have considered the most recent list (and relevant parameter errors) of VBs maintained by the LISA Consortium and available at~\citet{KupferTable} \citep[see also][]{2023arXiv230212719K}, and carried out the work within a Bayesian framework using the three independent TDI channels $A$, $E$, and $T$ and the instrument performance according the current science requirements. 

We have shown that \LISA{} will detect 25 currently known systems over its nominal mission lifetime with SNR in the range $\approx 6 \text{--} 100$. We have also quantified the expected accuracy with which the system parameters will be measured by computing their marginalised posterior probability distributions, and shown that \LISA{} will provide astrophysically non-trivial measurements for quantities such as orbital inclination and the evolution of the orbital period. We have also quantified the extent to which prior knowledge of the source parameters affects the integration time required to detect a VB, therefore providing an early ``verification'' of \LISA{}'s performance.

Consistent with previous studies, we have derived these baseline results under the assumption that the noise affecting the measurements is known and that within the frequency band covered by a VB signal no other GW source is present. Both assumptions are clearly wrong, and one may wonder how reliable these results, as well as all of those present in the literature, actually are.
We have therefore generalised our analysis in the two key directions that remove these oversimplifications. 

First, we have relaxed the assumption that the noise, both coming from the instrument and from the unresolved foreground of Galactic and extragalactic UCBs, is known in advance. By including the noise level (parameterised by a single parameter in each TDI channel) as one of the parameters that need to be fitted in the analysis, we have shown that there is minimal effect on the accuracy with which the VB parameters can be measured. As a by-product of this analysis, we have demonstrated that the overall PSD in each of the three channels can be measured to $\sim 8\%$, but we stress that this measurement is dependant on the number of frequency bins included in the analysis. 

Secondly, we have accounted for the fact that the signal from VBs will overlap with those from many other UCBs in the Galaxy. By considering a state-of-the-art synthetic population of Galactic binaries, we have analysed a small ($\approx 1000\,\mathrm{nHz}$) frequency band fitting concurrently for a VB, an unknown number of other DWDs, and the noise level in the three TDI channels. We have shown that the baseline results presented in Table~\ref{tab:GW_table} are robust.

The actual analysis of the \LISA{} data to identify VBs will need to include additional refinements that we have not considered here. 
In particular, we have assumed that the noise contribution is Gaussian and stationary throughout the observing time, which we know is not true. For example, the (dominant, in the band of interest for this work) confusion noise level changes during the year due to the \LISA{} motion. We have assumed there are no data gaps (which surely will occur), and we have not considered either transients of instrumental or astrophysical nature (e.g.\ radiation from a loud massive black hole binary) in the frequency band.

\section*{Acknowledgements}

We thank all the developers of the \BALROG{} codesuite, including those who are not authors here.
We also thank Thomas Kupfer for maintaining an up-to-date table of \LISA{} VBs that has been adopted for our study.
Computational resources used for this work were provided by the University of Birmingham’s BlueBEAR High Performance Computing facility.
AK, HM, CJM, and AV acknowledge the support of the UK Space Agency, Grant No. ST/V002813/1.
AV acknowledges the support of the Royal Society and Wolfson Foundation.

\section*{Data Availability}
 
A data release with all posterior samples, SNR calculations, and code to reproduce figures is available at \citet{finch_eliot_2022_7211349}.


\bibliographystyle{mnras}
\bibliography{refs}


\appendix

\section{Noise curves and characteristic strain}\label{sec:noisecurves}

This appendix gives details of the \LISA{} instrumental and confusion noise sources.
This appendix also gives useful equations for predicting the SNR of DWD sources computed in a low-frequency approximation. These equations were used to produce the results in the top panel of Fig.~\ref{fig:CharacteristicStrain}.
We stress that elsewhere in the paper all SNRs were computed using the full \LISA{} TDI outputs described in the main text, \emph{without} making a low-frequency approximation.

In the top panel of Fig.~\ref{fig:CharacteristicStrain}, the instrumental noise curve is plotted as $s_{\rm inst}(f) = \sqrt{f S_{\rm inst}(f)}$, with $S_{\rm inst}(f)$ being the low-frequency approximation of the PSD in line with the latest SciRD document~\citep{2021arXiv210801167B}:
\begin{equation}
    S_{\rm inst}(f) = (4S_{\rm disp}(f) + S_{\rm opt}(f))\left(1+0.6\left(\frac{2\pi f 2.5\times 10^9}{c}\right)^2\right),
\end{equation}
with $S_{\rm disp}(f)$ being the displacement noise and $S_{\rm opt}(f)$ being the optical noise.

The total noise curve is plotted as $s_n(f) = \sqrt{f S_n(f)}$, where $S_n(f) = S_{\rm inst}(f) + S_{\rm conf}(f)$, and the expression for the confusion noise PSD $S_{\rm conf}(f)$ is obtained from \citet{Babak:2017tow}:
\begin{align}\label{eq:conf_psd}
    S_{\rm conf}(f) = & A_{\rm gal} \left( \frac{f}{1 \,\mathrm{Hz}} \right) ^{-7/3} \exp{\left[ -\left(\frac{f}{s_1}\right)^\alpha\right]} \nonumber \\
    & \times \frac{1}{2}\left[1+\tanh{\left(-\frac{f-f_0}{s_2}\right)}\right],
\end{align}
where the parameters $A_{\rm gal}$, $\alpha$, $s_1$, $f_0$, and $s_2$ where fitted for different values of the observation time (Babak, private communication) and then interpolated as a function of $T_{\rm obs}$. 
Note that an updated confusion noise curve can be found in~\citep{Karnesis:2021tsh}.

The binaries considered here are nearly monochromatic, and emit at low frequencies with respect to the instrumental transfer frequency $f^* = c/2 \pi L \approx 19\,\mathrm{mHz}$, where $L$ is the length of the \LISA{} laser arms.
Therefore, for the sensitivity estimates used in the top panel of Fig.~\ref{fig:CharacteristicStrain}, the signal is modelled in the low-frequency approximation~\citep{Cutler:1997ta}. The combined signal is equivalent to two independent detectors, with
\begin{align}
    h_{I,II}(t) &= \frac{\sqrt{3}}{2} \left[F_+^{I,II}(t) h_+(t) + F_\times^{I,II}(t) h_\times(t) \right], \\
    h_+(t) &= -\mathcal{A}  (1 + \cos^2 \iota) \cos [2 \pi f_0 (t - \bm{\hat{k}} \cdot \bm{P})  + \phi_0], \\
    h_\times(t) &= 2 \mathcal{A} \cos \iota \sin [2 \pi f_0 (t - \bm{\hat{k}} \cdot \bm{P})  + \phi_0],
\end{align}
where $F_{+,\times}^{I,II}(t)$ are the detector antenna pattern functions and $c \bm{P}$ is the position of \LISA{}'s barycentre. The parameter $f_0$ is the GW frequency of the signal, $\phi_0$ is its initial phase, $\iota$ is its inclination, $\bm{\hat{k}}$ is the wave propagation vector, and $\mathcal{A}$ is the GW strain amplitude given by Eq.~\eqref{eq:amp}. 
The combined squares SNR $\rho^2$ for this signal is
\begin{align}
    \rho^2 &= 4 \int_0^\infty \frac{|\tilde{h}_I(f)|^2 + |\tilde{h}_{II}(f)|^2}{S_n(f)} \mathrm{d}f. \label{eq:SqSNR}
\end{align}

We can model the antenna pattern functions in the following way~\citep{Cutler:1997ta}:
\begin{align}
F_+^I(t) &= \frac{1}{2} \left( 1 + \cos^2 \bar{\theta} \right) \cos 2\bar{\phi} \cos 2 \bar{\psi} -  \cos\bar{\theta} \sin 2 \bar{\phi} \sin 2 \bar{\psi}, \\
F_\times^I(t) &= \frac{1}{2} \left( 1 + \cos^2 \bar{\theta} \right) \cos 2\bar{\phi} \sin 2 \bar{\psi} + \cos\bar{\theta} \sin 2 \bar{\phi} \cos 2 \bar{\psi}, \\
F_+^{II}(t) &= \frac{1}{2} \left( 1 + \cos^2 \bar{\theta} \right) \sin 2\bar{\phi} \cos 2 \bar{\psi} +  \cos\bar{\theta} \cos 2 \bar{\phi} \sin 2 \bar{\psi}, \\
F_\times^{II}(t) &= \frac{1}{2} \left( 1 + \cos^2 \bar{\theta} \right) \sin 2\bar{\phi} \sin 2 \bar{\psi} -  \cos\bar{\theta} \cos 2 \bar{\phi} \cos 2 \bar{\psi}, 
\end{align}
where $\bar{\theta}$ and $\bar{\phi}$ are the spherical angles of the source position in a frame rotating together with \LISA{}'s arms, and $\bar{\psi}$ is a polarisation angle. We can find $\bar{\theta}$ and $\bar{\phi}$ as:
\begin{align}
    \cos \bar{\theta} &= \bm{\hat{N}} \cdot \bm{\hat{\bar{z}}}, \\
    \tan \bar{\phi} &= \frac{\bm{\hat{N}} \cdot \bm{\hat{\bar{y}}}}{\bm{\hat{N}} \cdot \bm{\hat{\bar{x}}}},
\end{align}
where $\bm{\hat{N}}$ is a unit vector pointing towards the source, and $\bm{\hat{\bar{x}}}$, $\bm{\hat{\bar{y}}}$, $\bm{\hat{\bar{z}}}$ form a triad defining the detector frame, given in an inertial frame tied to the ecliptic by
\begin{align}
    \bm{\hat{N}} &= ( \cos b \cos l, \cos b \sin l, \sin b), \\
    \bm{\hat{\bar{x}}} &= \left( \frac{1}{4} \left( 3 - \cos 2 \omega t \right) , - \frac{1}{4} \sin 2 \omega t , \frac{\sqrt{3}}{2} \cos \omega t \right), \\
    \bm{\hat{\bar{y}}} &= \left(- \frac{1}{4} \sin 2 \omega t , \frac{1}{4} \left( 3 + \cos 2 \omega t \right) , \frac{\sqrt{3}}{2} \sin \omega t \right), \\
    \bm{\hat{\bar{z}}} &= \left(-\frac{\sqrt{3}}{2} \cos \omega t , -\frac{\sqrt{3}}{2} \sin \omega t  , \frac{1}{2} \right),
\end{align}
where $b$ and $l$ are respectively the ecliptic latitude and longitude of the source, and $\omega = 2\pi/{\rm yr}$ is the orbital angular frequency of the \LISA{} constellation around the Sun.

In order to compute the Fourier transform of the signal, we can take advantage of the fact that the GW frequency $f_0$ is well separated from the frequency of the modulation from \LISA{}'s motion and compute it in a time interval $1/f_0 \ll \Delta T \ll 2\pi/\omega$ . We can write
\begin{align}
    \tilde{h}_{I,II} (f) &= \sum_{n = 0}^{T_{\rm obs}/\Delta T -1} \tilde{h}_{I,II} (f, n), \\
    \tilde{h}_{I,II} (f, n) &= \int_{n \Delta T}^{(n+1) \Delta T} h_{I,II}(t) e^{-2\pi i f t} \mathrm{d}t \\
    &\approx \frac{\sqrt{3}}{2} B_+^{I,II}(n \Delta T) \int_{n \Delta T}^{(n+1) \Delta T} h_+(t) e^{-2\pi i f t} \mathrm{d}t \nonumber\\
    &+ \frac{\sqrt{3}}{2} B_\times^{I,II}(n \Delta T) \int_{n \Delta T}^{(n+1) \Delta T} h_\times(t) e^{-2\pi i f t} \mathrm{d}t \nonumber\\
    &= \frac{\sqrt{3}}{4} [ A_+ B_+^{I,II}(n \Delta T) - i A_\times B_\times^{I,II}(n \Delta T) ] \nonumber\\
    &\times \left\{ g[(n+1) \Delta T] - g(n \Delta T) \right\}, \\
    B_+^{I,II}(t) &= F_+^{I,II}(t) e^{- 2\pi i f_0\ \bm{\hat{k}} \cdot \bm{P}(t)},\\
    B_\times^{I,II}(t) &= F_\times^{I,II}(t) e^{2\pi i f_0\ \bm{\hat{k}} \cdot \bm{P}(t)}, \\
    g(t) &= \frac{\sin [\pi (f - f_0) t]}{\pi (f - f_0)} e^{- i \pi (f - f_0) t + i \phi_0} \nonumber\\
    &+ \frac{\sin [\pi (f + f_0) t]}{\pi (f + f_0)} e^{- i \pi (f + f_0) t - i \phi_0}.
\end{align}
Note that since the square SNR in Eq.~\eqref{eq:SqSNR} is obtained by integrating over positive frequencies and since this Fourier transform has narrow support, we can neglect the second line in $g(t)$.

Since the pre-factor varies slowly, if the observation window comprises an integer number of years, we can average the pre-factor in this result and obtain the approximation
\begin{align}
    \left\langle |\tilde{h}_{I,II}(f)|^2 \right\rangle_t &\approx \frac{3 \mathcal{I}^{I,II}}{16} \frac{\sin^2[\pi (f - f_0) T_{\rm obs}]}{\pi^2 (f - f_0)^2}, \\
    \mathcal{I}^{I,II} &= \int_0^{T_{\rm obs}} \Big( A_+^2 F_+^{I,II}(t)^2 + A_\times^2 F_\times^{I,II}(t)^2 \\
    &+ 2 A_+ A_\times F_+^{I,II}(t) F_\times^{I,II}(t) \sin[4 \pi f_0\ \bm{\hat{k}} \cdot \bm{P}(t) ] \Big) \mathrm{d}t, \nonumber
\end{align}
where we assumed that the pre-factor in $\tilde{h}_{I,II}(f)$ is a constant equal to its root modulus square average, and simplified the factors of $g(t)$ by taking advantage of the fact that it is then a telescoping sum. 
We can further average over the polarisation angle $\bar{\psi}$ to get
\begin{align}
    \left\langle \rho^2 \right\rangle_{t,\bar{\psi}} &= 4 \int_0^\infty \frac{\left\langle |\tilde{h}_I(f)|^2 + |\tilde{h}_{II}(f)|^2 \right\rangle_{t,\bar{\psi}}}{S_n(f)} \mathrm{d}f.
\end{align}

The support of $g(T_{\rm obs})$ as a function of frequency is of the order of a few $\Delta f = 1/T_{\rm obs}$. Assuming that the noise PSD $S_n(f)$ is constant inside a window of a width of a few $\Delta f$ around $f_0$, we can compute:
\begin{align}
    \left\langle \rho^2 \right\rangle_{t,\bar{\psi}} &\approx \frac{3 \mathcal{A}^2 T_{\rm obs}}{4096\ S_n(f_0)} \left(443 - 78 \sin^2 b - 37 \sin^4 b\right) \nonumber\\
    &\times \left(1+ 6\cos^2 \iota + \cos^4 \iota \right).
\end{align}
We can similarly compute the SNR averaged over ecliptic latitude and/or inclination. We find
\begin{align}
    \left\langle \rho^2 \right\rangle_{t,\bar{\psi},\iota} &\approx \frac{3 \mathcal{A}^2 T_{\rm obs}}{1280\ S_n(f_0)} \left(443 - 78 \sin^2 b - 37 \sin^4 b\right), \\
    \left\langle \rho^2 \right\rangle_{t,\bar{\psi}, b} &\approx \frac{3 \mathcal{A}^2 T_{\rm obs}}{10\ S_n(f_0)} \left(1+ 6\cos^2 \iota + \cos^4 \iota \right), \\
    \left\langle \rho^2 \right\rangle_{t,\bar{\psi}, b , \iota} &\approx \frac{24 \mathcal{A}^2 T_{\rm obs}}{25\ S_n(f_0)}.
\end{align}

In order to represent the SNR of each VB as a ratio $h_c/s_n$, the characteristic strain is evaluated with the following formula:
\begin{align}
    \label{eq:CharStrain}
    &h_c = \\
    &\mathcal{A}\left[\frac{3 f_0 T_{\rm obs}}{4096}(443-78\sin^2{b}-37\sin^4{b})(1+6\cos^2{\iota}+\cos^4{\iota})\right]^{1/2}, \nonumber
\end{align}
where $T_\mathrm{obs} = 4\,\mathrm{yr}$ and [$f_0, b, \iota$] are the EM measurements of frequency, ecliptic latitude and inclination of the VB respectively (see Table~\ref{tab:EM_table}).
The error bars in the top panel of Fig.~\ref{fig:CharacteristicStrain} on each $h_\mathrm{c}$ are the result of the evaluation of the minimum and maximum characteristic strain, obtained by replacing [$f_\mathrm{min}, D_{L\mathrm{max}}, m_{1\mathrm{min}}, m_{2\mathrm{min}}, f(\iota)_\mathrm{max}$] and [$f_\mathrm{max}, D_\mathrm{min}, m_{1\mathrm{max}}, m_{2\mathrm{max}}, f(\iota)_\mathrm{min}$] respectively into Eq.~\eqref{eq:CharStrain}. $f(\iota)$ refers to the expression $f(\iota) = 1+6\cos^2{\iota}+\cos^4{\iota}$.
The minimum and maximum values of each parameter are reported in Table~\ref{tab:EM_table}.

\bsp	
\label{lastpage}
\end{document}